\documentclass[%
 reprint,
superscriptaddress,
showpacs,preprintnumbers,
nofootinbib,
 amsmath,amssymb, 
 aps,
 prl,
 longbibliography,
]{revtex4-1}

\usepackage{cancel}
\usepackage{accents}
\usepackage{mciteplus,slashed}
\usepackage{amssymb,cancel,amsmath}
\usepackage{dcolumn}
\usepackage{bm}
\usepackage{soul}
\usepackage[caption=false]{subfig}
\usepackage{appendix}
\usepackage{physics}
\usepackage{booktabs}
\usepackage{feynmp-auto}
\usepackage[T1]{fontenc}	
\usepackage{csvsimple}
\usepackage[bookmarksnumbered=true]{hyperref} 
\hypersetup{
    colorlinks = true,
    linkcolor = blue,
    anchorcolor = blue,
    citecolor = blue,
    filecolor = blue,
    urlcolor = blue
    }
\usepackage[section]{placeins}
\usepackage[capitalise]{cleveref}
\usepackage{booktabs}
\usepackage{graphicx}
\usepackage{mathrsfs}
\graphicspath{{Figures/}}

\AtBeginDocument{
\heavyrulewidth=.08em
\lightrulewidth=.05em
\cmidrulewidth=.03em
\belowrulesep=.65ex
\belowbottomsep=0pt
\aboverulesep=.4ex
\abovetopsep=0pt
\cmidrulesep=\doublerulesep
\cmidrulekern=.5em
\defaultaddspace=.5em
}
\usepackage[dvipsnames]{xcolor}
\usepackage[normalem]{ulem}
\usepackage{fontawesome} 

\newcommand{\hyphen}{\,\mathchar`-\mathchar`-\,}

\definecolor{forestgreen}{HTML}{228B22}

\begin{document}

\title{Searching for axions with kaon decay at rest}

\author{Yohei Ema}
\email{ema00001@umn.edu}
\affiliation{William I. Fine Theoretical Physics Institute,
University of Minnesota, Minneapolis, MN 55455, USA}
\affiliation{School of Physics and Astronomy, University of Minnesota, Minneapolis, MN 55455, USA}

\author{Zhen Liu}
\email{zliuphys@umn.edu}
\thanks{\href{https://orcid.org/0000-0002-3143-1976}{0000-0002-3143-1976}}
\affiliation{School of Physics and Astronomy, University of Minnesota, Minneapolis, MN 55455, USA}

\author{Ryan Plestid}
\email{rplestid@caltech.edu}
\affiliation{Walter Burke Institute for Theoretical Physics, California Institute of Technology, Pasadena, CA 91125, USA}

\preprint{UMN-TH-4221/23}
\preprint{FTPI-MINN-23-13}
\preprint{CALT-TH/2023-028}

\begin{abstract}
    We describe a novel search strategy for axions (or hadronically coupled axion-like particles) in the mass range of $m_a \lesssim 350\,{\rm MeV}$. The search relies on kaon decay at rest, which produces a mono-energetic signal in a large volume detector (e.g.\ a tank of liquid scintillator) from axion decays $a\rightarrow \gamma\gamma$ or $a\rightarrow e^+e^-$.  The decay modes $K^+\to \pi^+ a$ and $a \to \gamma \gamma$ are induced by the axion's coupling to gluons, which is generic to any model which addresses the strong CP problem.     We recast a recent search from MicroBooNE for $e^+e^-$ pairs, and study prospects at JSNS$^2$ and other near-term facilities. We find that JSNS$^2$ will have world-leading sensitivity to hadronically coupled axions in the mass range of 
$40\,\mathrm{MeV} \lesssim m_a \lesssim 350\,\mathrm{MeV}$. 
\end{abstract}

 \maketitle

\textbf{Introduction:} 
The neutron's electron dipole moment (EDM), $d_n \lesssim 2 \times 10^{-26}~e~{\rm cm}$ \cite{Graner:2016ses,Abel:2020gbr}, is ten orders of magnitude smaller than its naive estimate of $5\times 10^{-16} e~{\rm cm}$. This is unexpected since charge-parity (CP) is violated within the Standard Model (SM), and therefore no fundamental symmetry forbids a neutron EDM.  This is the strong CP problem, and its  microscopic origin can be traced to the minuscule value of the QCD $\bar{\theta}$ parameter, $\bar{\theta}\lesssim 10^{-10}$, which controls the unique CP-violating QCD coupling in the SM. Axions are a popular solution that provide a dynamical mechanism for the relaxation of $\bar{\theta}$ \cite{Peccei:1977hh,Peccei:1977ur,Weinberg:1977ma,Wilczek:1977pj};  if the axion's potential is generated exclusively by QCD, it naturally aligns the axion's ground state with $\bar{\theta}=0$. 

Axions generically suffer from the so-called quality problem \cite{Kamionkowski:1992mf,Barr:1992qq,Ghigna:1992iv,Holman:1992us} wherein high energy contributions to the axion potential can (and often do) displace the minimum away from $\bar{\theta}=0$. These problem does not arise if the QCD potential is ``strenghtened'', for instance by introducing a mirror QCD sector. These non-minimal axion models produce a potential that is robust against the high energy corrections discussed above and often predict heavier axions as compared to minimal  axion models~\cite{Dimopoulos:1979pp,Tye:1981zy, Holdom:1982ex, Flynn:1987rs, Rubakov:1997vp,Berezhiani:2000gh,Hook:2014cda,Fukuda:2015ana,Gherghetta:2016fhp, Dimopoulos:2016lvn, Agrawal:2017ksf, Agrawal:2017evu, Gaillard:2018xgk, Lillard:2018fdt, Fuentes-Martin:2019bue, Csaki:2019vte, Hook:2019qoh,Gherghetta:2020keg, Gherghetta:2020ofz, Valenti:2022tsc,Kivel:2022emq}.

Recent investigations have reignited interest in these so-called heavy axion models with $m_a\gtrsim 1\,{\rm MeV}$. More generally, there is a broad interest in axion like particles (ALPs) \cite{Bauer:2017ris,Irastorza:2018dyq} which may serve as IR messengers of UV completions such as a string landscape \cite{Arvanitaki:2009fg,Jaeckel:2010ni}. A wide range of search strategies have been proposed ranging from beam dumps, to flavor facilities, and collider experiments~\cite{Essig:2010gu, Dobrich:2015jyk, Dolan:2017osp, Dobrich:2019dxc, Harland-Lang:2019zur, Dent:2019ueq, Brdar:2020dpr, Jaeckel:2012yz, Mimasu:2014nea, Jaeckel:2015jla, Izaguirre:2016dfi, Knapen:2016moh, Bauer:2017ris, Brivio:2017ije, Mariotti:2017vtv, CidVidal:2018blh, Beacham:2019nyx, Alonso-Alvarez:2018irt, Ebadi:2019gij, Gavela:2019cmq, Gavela:2019cmq, Altmannshofer:2019yji, Gershtein:2020mwi, DiLuzio:2020wdo, Knapen:2021elo,Co:2022bqq,ArgoNeuT:2022mrm}. 

In this work, we point out that kaon decay at rest (KDAR) offers a powerful probe of axions (or hadronically coupled ALPs) in the mass range of $m_a=40 \hyphen 350~{\rm MeV}$. Unlike previous studies of  KDAR \cite{Coloma:2022hlv,Batell:2019nwo} the axions we consider are naturally coupled to quarks and/or gluons such that the hadronic decays of kaons serve as a powerful axion factory. The signal is a visible decay of either $a\rightarrow \gamma\gamma$ or $a\rightarrow e^+e^-$. The dominant SM kaon decay modes are  $K^+ \rightarrow \mu^+ \nu_\mu$, as well as $K^+\rightarrow \pi^0 e^+ \nu_e$,\!\footnote{Negatively charged $K^-$ decays are subdominant to $K^-$ capture on nuclei. This cannot occur for $K^+$ because it does not form bound states with nuclei.} and both channels produce a neutrino flux that can be measured~\cite{Spitz:2014hwa,Nikolakopoulos:2020alk}.  In addition to predictions of hadronic cascade simulations, this provides an experiment with an {\it in situ} measurement of their KDAR population. The axion rate is then calculable, and a counting experiment can be performed.

For concreteness, we consider an axion coupled to Standard Model (SM) gauge bosons via
\begin{equation}
    \begin{split}
	\mathcal{L}_a &= \frac{1}{2}(\partial a)^2 - \frac{m_a^2}{2}a^2
	+ c_{GG}\frac{\alpha_s}{4\pi}\frac{a}{f}G_{\mu\nu} \tilde{G}^{\mu\nu}
	    \\
	&+ c_{WW}\frac{\alpha_2}{4\pi}\frac{a}{f}W_{\mu\nu}\tilde{W}^{\mu\nu}
	+ c_{BB}\frac{\alpha_Y}{4\pi}\frac{a}{f}B_{\mu\nu}\tilde{B}^{\mu\nu},
 \end{split}
\end{equation}
at a high scale, where $a$ is the axion, and $\alpha_s = g_s^2/4\pi$, $\alpha_2 = g^2/4\pi$ and $\alpha_Y = {g'}^2/4\pi$. The coupling $c_{GG}$ can be absorbed into $f$ and we define the axion decay constant as $f_a = f/2c_{GG}$. As two useful benchmarks, we focus on the so-called ``gluon dominance'' ($c_{GG} \neq 0$, $c_{WW} = c_{BB} = 0$), and ``co-dominance'' ($c_{GG} = c_{WW} = c_{BB} \neq 0$)  scenarios following~\cite{Kelly:2020dda,Jerhot:2022chi,Afik:2023mhj} (using the relative normalization of $c_{BB}$ in~\cite{Jerhot:2022chi,Afik:2023mhj}). We do not consider, e.g.\ axion couplings to quarks, though their inclusion is straightforward. In what follows, we discuss the theory of $K^+\rightarrow \pi^+ a$, and project sensitivities for JSNS$^2$ and a MicroBooNE $\gamma\gamma$ search.


\textbf{Kaon decay at rest:} 
In models with a hadronically coupled axion, $K^+\rightarrow \pi^+ a$ serves as a powerful axion production channel. Since axions are long-lived, they decay in flight to visible final states. The resulting signal is a mono-energetic peak at an energy of 
\begin{equation}
    E_a =  \frac{m_K^2 + m_a^2 - m_\pi^2}{2 m_K} ~.
    \label{eq:axion_energy}
\end{equation}
This indicates that the heavier axion carries more energy and its minimal value is given by setting 
$m_a = 0$ as $E_a > 227\,\mathrm{MeV}$.
The branching ratio for $K^+ \rightarrow \pi^+ a$ can be reliably predicted in chiral perturbation theory \cite{Bauer:2021wjo}, with the result [neglecting terms of $\mathcal{O}(m_a^2/m_K^2)$  and $\mathcal{O}(m_\pi^2/m_K^2)$]
\begin{align}
	\mathrm{BR}(K^+\to \pi^+ a) &= \frac{\tau_{K^-}}{\tau_{K_S}} \frac{f_\pi^2}{8 f_a^2}
	\times \mathrm{BR}(K_S \to \pi^+ \pi^-)~.
	\label{BR-eq}
\end{align}
\Cref{BR-eq} receives corrections from finite axion and pion mass effects in both the matrix element and phase space, both of which are taken into account in our numerical estimates (see~\cite{Bauer:2021wjo} for the complete expression). In the co-dominance case, the axion coupling to the $W$-boson induces an extra contribution~\cite{Izaguirre:2016dfi}.
However, this contribution is negligible for $c_{WW} = c_{GG}$ and we do not consider it further in what follows.

\begin{figure}[t]
   \includegraphics[width=\linewidth]{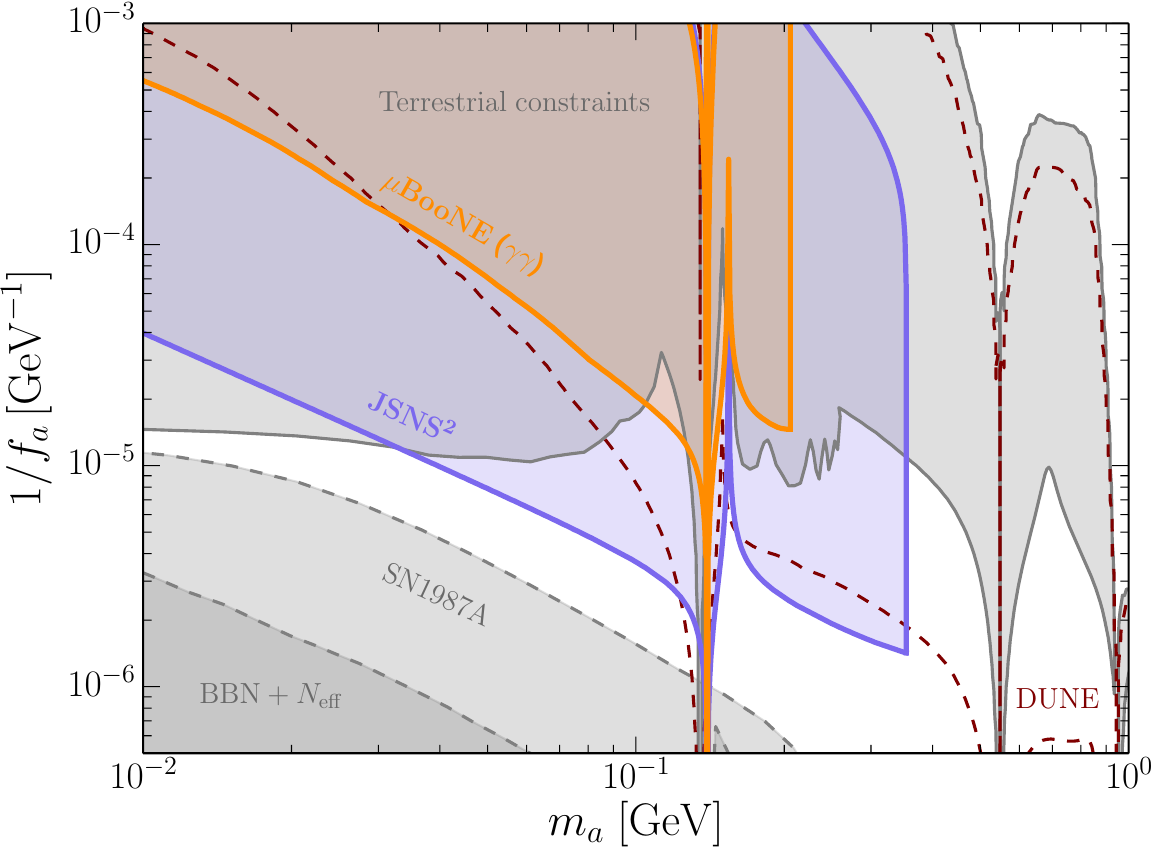}
    \caption{\label{fig:gluon} (\textbf{Gluon Dominance}) Sensitivities of MicroBooNE 
    and JSNS$^2$ compared with existing limits and other projected sensitivities
    when all couplings are induced by a gluon coupling $c_{GG}$ at a high scale. 
    The MicroBooNE sensitivity is cut at 210 MeV because that is the range that appears in \cite{MicroBooNE:2021usw}. 
    Existing limits include constraints from SN1987A~\cite{Chang:2018rso,Ertas:2020xcc} and cosmology~\cite{Depta:2020wmr} adapted from~\cite{Kelly:2020dda}, 
    Kaon decays (E949~\cite{BNL-E949:2009dza} and NA62~\cite{NA62:2021zjw}),
    and beam dump searches (CHARM~\cite{CHARM:1985anb} and NuCal~\cite{Blumlein:1990ay}), with data adapted from~\cite{Jerhot:2022chi}.
    We also show projected sensitivity of DUNE~\cite{Kelly:2020dda}.
    Other projected sensitivities not shown here include DarkQuest~\cite{SeaQuest:2017kjt,Berlin:2018pwi,Blinov:2021say}, FASER~\cite{FASER:2018eoc}, KOTO~\cite{Yamanaka:2012yma,Afik:2023mhj}, and SHiP~\cite{SHiP:2015vad}.
    We have re-derived constraints from E949 and NA62 and our results agree with~\cite{Afik:2023mhj} (but disagree with~\cite{Ertas:2020xcc} and therefore the curves in~\cite{Kelly:2020dda}). 
    }
\end{figure}

Axions produced from KDAR can decay inside nearby detectors, either to $\gamma\gamma$ or $e^+ e^-$. The number of axions produced at a KDAR source is given by
\begin{equation}
    N_a= \frac{{\rm BR}(K^+\rightarrow \pi^+ a)~}{ {\rm BR}(K^+\rightarrow  \mu^+ \nu_\mu )} N_{\nu_\mu}~,
\end{equation}
where $N_{\nu_\mu}$ is the number of muon neutrinos from KDAR that are produced in the beam stop. 
In the limit of a long decay length, $\lambda_a \gg L$ where $L$ is the distance from the KDAR source to the detector, the number of axions that decay in the detector is given by\footnote{
	In our numerical computation, we do not rely on this approximation.
    Instead, we include the finite axion decay length properly to obtain the upper limit of the sensitivity correctly.
}
\begin{equation}
    \label{a-rate}
    N_{a\rightarrow {\rm vis} } = N_a\times \frac1{4\pi L^2}\frac{V}{\lambda_a(m_a)}~,
\end{equation}
where $V$ is the volume of the detector, $\lambda_a= \beta_a \gamma_a \tau_a$ is the decay length of the axion in the lab frame, and we have assumed all final states of the axion decay are visible.
Our energy range of interest is given by
\begin{align}
	227\,\mathrm{MeV} < E_{ee}, E_{\gamma\gamma} < 354\,\mathrm{MeV},
\end{align}
where the lower bound (upper bound) is given by setting $m_a = 0~(m_a = m_{K^+} - m_{\pi^+})$ in \Cref{eq:axion_energy}.

Assuming $a\rightarrow \gamma\gamma$ dominates over $a\rightarrow e e$, which is generically true for   theories without direct axion-electron couplings, the lifetime of the axion is given by\footnote{In models with a mirror sector the axion may have invisible decay modes with $O(1)$ branching ratios.}
\begin{equation}
    \tau_a^{-1} = \frac{\alpha^2 m_a^3}{256 \pi^3 f^2_a} \abs{c_{\gamma\gamma}^{\rm eff}}^2~,
\end{equation}
where $\alpha$ is the electromagnetic fine structure constant. The effective coupling to the photon is given by~\cite{Bauer:2017ris,Blinov:2021say}
\begin{equation}    
    \label{photon-coupling}
    c_{\gamma\gamma}^{\rm eff} \approx c_{\gamma\gamma}
    -\qty(\frac53 + \frac{m_\pi^2}{m_\pi^2-m_a^2} \frac{m_d-m_u}{m_u+m_d}) ~, 
\end{equation}
where $c_{\gamma\gamma} = 0$ in the gluon dominance case and $c_{\gamma\gamma} = 2$ in the co-dominance case,
and our constraints are expressible in terms of $1/f_a^4$.
\Cref{photon-coupling} assumes a two-flavor approximation, which is reasonably accurate for the masses we consider, $m_a\lesssim 350~{\rm MeV}$. For these masses we never approach regions of resonant mixing with $\eta$ and $\eta'$; more complete three flavor expressions can be found in Appendix B of \cite{Blinov:2021say}.  The constraints can be easily re-scaled for other model-dependent choices (e.g.\ with the axion couplings to the up and down quarks). It is also straightforward to project sensitivities for models where $a\rightarrow e^+e^-$ is the dominant decay channel. 

For $m_a$ close to $m_\pi$ cancellations can occur such that $c_{\gamma\gamma}^{\rm eff}$ vanishes.\footnote{
This is due to a cancellation among the axion direct coupling to the photon and the axion-pion mass and kinetic mixing contributions.
Somewhat related to this, \Cref{photon-coupling} relies on the small mixing angle expansion, and hence is applicable only when 
	$\vert m_\pi^2/(m_\pi^2 - m_a^2) \times f_\pi/f\vert \ll 1$, strictly speaking.
	Since this condition is violated only for the axion mass very close to $m_\pi$, we ignore this subtlety.
}
For our choice of the quark mass ratio, $m_u/m_d = 0.46$, 
this happens only in the gluon dominance case. 
We note however that if we instead use $m_u/m_d = 0.56$ motivated from the meson mass spectrum, 
the cancellation happens around $m_a \simeq 55\,\mathrm{MeV}$ in the co-dominance case.
If the cancellation happens the decay length of the axion is set by $a\rightarrow e^+e^-$. Such a coupling is always generated by radiative effects. For our numerical estimates, we include the $a\rightarrow e^+e^-$ decay channel, taking $g_{aee}=0.37\times 10^{-3}~c_{GG}$ (see Eq.~(62) of \cite{Bauer:2020jbp}), 
although the effect of this coupling is almost invisible in our plots. 
Substantially stronger coupling to electrons is possible if it arises at tree-level in the UV or is induced by top-quarks \cite{Bauer:2020jbp}; in these cases our constraints would strengthen. In principle for $m_a \geq 2 m_\mu$ the muon decay channel can also be included, but this only affects the ceiling of our constraint by an $\mathcal{O}(1)$ factor since $\Gamma(a\rightarrow \gamma\gamma)$ is comparable to $\Gamma(a\rightarrow \mu^+\mu^-)$. If a detector can measure muon tracks, then $a\rightarrow \mu^+ \mu^-$ would present an additional signal channel with much lower backgrounds.  However, since the impact of muon couplings on our results is relatively modest for the co-dominance and gluon dominance scenarios, and this region can be probed by other experiments~\cite{Co:2022bqq,ArgoNeuT:2022mrm}, we do not discuss it further. 

\textbf{MicroBooNE search for heavy scalars:} 
Using the above formulae, we can recast a recent search by MicroBooNE \cite{MicroBooNE:2021usw}. Their search channel was $K\rightarrow \pi h_D$ followed by $h_D\rightarrow e^+ e^-$ with $h_D$ a dark Higgs \cite{Batell:2019nwo}. If we consider the $a\rightarrow e^+e^-$ search, then mapping their result to a heavy axion is an  immediate constraint on $g_{aee}$. Using the induced $g_{aee}$ from gluon couplings mentioned above, we then obtain a limit on $f_a$. We find that the constraints on $f_a$ obtained in this gluon-dominance scenario are very weak, excluding $f_a\lesssim 1~{\rm TeV}$, which is already ruled out by other experiments, and so we do not include this in our summary plot
(see~\cite{Coloma:2022hlv} for ALPs coupled exclusively to electroweak bosons).

For the $a\rightarrow \gamma \gamma$ channel, it is not possible to interpret the MicroBooNE result as a constraint.  The search in \cite{MicroBooNE:2021usw} made use of a boosted decision tree  (BDT) in classifying their events and should reject $\gamma\gamma$ topologies.\!
\footnote{In principle there should be some probability of a $\gamma\gamma$ pair contaminating the BDT-tagged $e^+e^-$ sample, but this would require collaboration input.}
As a crude estimate of the sensitivity, we may assume that a dedicated analysis for $\gamma\gamma$ final states is performed with comparable BDT performance. Then the sensitivity to $f_a$ (from the same dataset) would be given by equating
$\Gamma(K \to \pi h_D) \times \Gamma(h_D \to e e)$, evaluated
with the upper bound on the mixing angle in~\cite{MicroBooNE:2021usw},
to our $\Gamma(K \to \pi a) \times \Gamma(a \to \gamma\gamma) \times e^{-L/\lambda_a}$
with $L \simeq 100\,\mathrm{m}$ the distance between MicroBooNE and the NuMI absorber (i.e.\ the KDAR source). This treatment is valid since the decay length of the dark Higgs is much longer than $L$ for the mixing angles in~\cite{MicroBooNE:2021usw},
and because the MicroBooNE detector is much smaller than $L$.

\begin{figure}[t]
    \includegraphics[width=\linewidth]{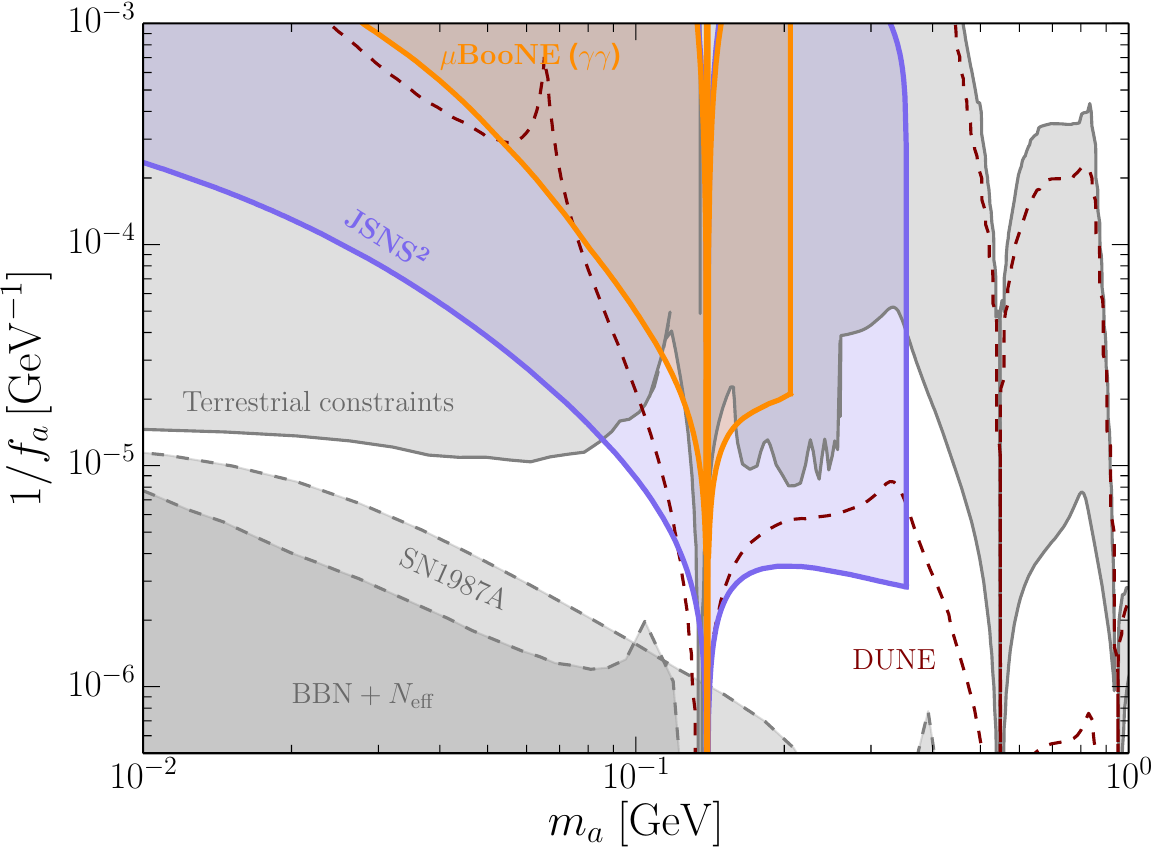}
    \caption{(\textbf{Co-Dominance}) 
    The same figure as \Cref{fig:gluon} but with $c_{WW} = c_{BB} = c_{GG}$.
    The sensitivity for $m_a \ll m_\pi$ is worse than the gluon dominance case 
    since $\vert c_{\gamma\gamma}^{\mathrm{eff}} \vert \ll 1$ for this specific choice of the parameters.
    }
    \label{fig:co}
\end{figure}

In \Cref{fig:gluon} (gluon dominance) and \Cref{fig:co} (co-dominance), 
we plot the sensitivity of MicroBooNE estimated in this way by the orange lines.
 MicroBooNE may be able to explore certain small regions of parameter space not covered by existing experiments if a dedicated search for $\gamma\gamma$ final states is performed; this is qualitatively similar to the situation with a dark Higgs \cite{Batell:2019nwo,MicroBooNE:2021usw}. As alluded to above, searches for dimuon final states may also be of interest for $m_a\geq 2 m_\mu$ since MicroBooNE can easily reconstruct $\mu^+\mu^-$ pairs.

\textbf{Searches at JSNS$^2$:} 
Next, we consider JSNS$^2$ \cite{Ajimura:2017fld,JSNS2:2021hyk,Maruyama:2022juu}, which is designed to test the excess of events seen at LSND \cite{LSND:2001aii}. A crucial difference between JSNS$^2$ and LSND 
is the proton beam energy, $3~{\rm GeV}$ vs $0.8~{\rm GeV}$, such that JSNS$^2$ serves as both a $\pi$DAR and KDAR facility, whereas LSND had much fewer KDAR events (if any) \cite{Gudima:2009zz}. We, therefore, find that JSNS$^2$ offers much more compelling sensitivity to heavy axions and do not consider LSND further in what follows. 

To estimate the sensitivity at JSNS$^2$ we take the number of KDAR neutrinos per proton on target ($N_{\nu_\mu}= 0.0034/{\rm POT}$) from~\cite{Ajimura:2017fld}.  Although the kaon production rate is uncertain ($N_{\nu_\mu}$ can be twice as large~\cite{Axani:2015dha}), 
as we mentioned in the Introduction, in the actual experiment the KDAR neutrino flux can be measured \emph{in situ}.
Using estimates from Fig.\ 3 of \cite{Jordan:2018gcd}, we find that the search will be nearly background free $N_{\rm bkg}\leq 1$ in three years of operation,
except for  $E_a\leq 238~{\rm MeV}$ (corresponding to $m_a < 104\,\mathrm{MeV}$) 
where muon neutrinos produced from kaon decay-at-rest comes into play and
$N_{\rm bkg} \approx 2.5$ over three years. Ref.\  \cite{Jordan:2018gcd} assumes an additional lead shielding of the detector, which has not been put in place at JSNS$^2$. Cosmic backgrounds are therefore underestimated in that work. However, the axion signal we consider benefits from the high energy signal region, $E > 227\,\mathrm{MeV}$ (see \Cref{eq:axion_energy}), where cosmic events are suppressed. Further reductions in the cosmic muon rate can be achieved using coincident tagging with daughter Michel electrons \cite{hyoungku_jeon_2021_5784148}. Since studies of cosmic background mitigation are ongoing, 
we plot projections using a 5-event contour 
(corresponding to approximately $95$\% C.L. exclusion if JSNS$^2$ observes only one event) 
assuming 100\,\% efficiency of the detector.\footnote{
        The background rate is expected to be lower for a higher energy bin,
	and hence the statistical meaning of the 5-event contour varies depending on the axion mass
        (which determines the signal energy).
	However, as long as the background event rate is smaller than unity, this effect is minor.
}
Backgrounds will be smooth above $227~{\rm MeV}$, and a sideband analysis can be used to estimate their size {\it in situ} such that a search will always be statistics rather than systematics limited. As we discuss below, if backgrounds are high at JSNS$^2$ a search could be performed using their second detector~\cite{Ajimura:2020qni,Maruyama:2022juu}.

To compute the expected number of axion events, we take $L=24~{\rm m}$ as the distance between the KDAR source and the JSNS$^2$ detector, and the detector volume $V=(17~{\rm tonnes})/(0.852~{\rm g/ml})=20.35~{\rm m}^3$ \cite{JSNS2:2021hyk,kim2011measurement}. We assume $10^{23}$ POT, corresponding to roughly three years of live time. This results in $\sim 3\times 10^{20}$ stopped $K^+$ in total, which is much larger than the $10^{12} \hyphen 10^{13}$  $K^+$ decay events at NA62~\cite{NA62:2021zjw}.

In \Cref{fig:gluon} (gluon dominance) and \Cref{fig:co} (co-dominance), we plot the sensitivity of JSNS$^2$ by the blue lines, together with the existing constraints and future sensitivities. The figures demonstrate that JSNS$^2$ has excellent sensitivity to heavy axions. Note that the event number scales as $f_a^{-4}$. Therefore, even if we instead require e.g.\ 50 events, the sensitivity only weakens by  $10^{1/4} \simeq 1.8$, which does not alter our main conclusion.

In our analysis, we have focused on the JSNS$^{2}$ near detector.
However, the JSNS$^2$ collaboration recently installed another detector at a far location~\cite{Ajimura:2020qni,Maruyama:2022juu}, $48\,\mathrm{m}$ away from the source. This second detector is expected to start taking data soon and contains $32$ tonnes of liquid scintillator as a fiducial volume;  the larger volume partially compensating for the longer baseline. Together with possibly smaller backgrounds due to its location, the far detector may be better suited for axion searches once it starts its operation.

\textbf{Conclusions:}
KDAR provides a clean smoking gun signature of hadronically coupled axions. A $K^+$ production target, coupled with a large volume detector placed $\sim 10\text{--}100~{\rm m}$ away, allows for a powerful probe of visibly decaying particles lighter than the kaon (e.g.\ dark scalars or heavy neutral leptons \cite{Batell:2019nwo,Kelly:2021xbv}). In the context of KDAR, axions are particularly compelling due to their well-motivated hadronic couplings, which are necessary in any model that addresses the strong CP problem. 

We have focused on two benchmark scenarios (gluon dominance and co-dominance) for ease of comparison with the literature. It is interesting to understand how constraints vary with model-dependent coupling textures. The visible decays we consider here are governed both by ${\rm BR}(K\rightarrow \pi a)$ and the axion decay length and scale as $1/f_a^4$. By way of contrast, the constraints from NA62 depend only on ${\rm BR}(K\rightarrow \pi a)$ and scale as $1/f_a^2$. Stronger hadronic couplings will therefore favor NA62 over JSNS$^2$. Conversely, weaker hadronic couplings  and/or a larger $c_{\gamma\gamma}^{\rm eff}$ will favor JSNS$^2$ over NA62. Beam dump searches scale the same way as JSNS$^2$.

We find that JSNS$^2$ will have world-leading sensitivity to  heavy axions. One might imagine a competitive experimental landscape of modern high-intensity low-energy proton beams.  Notably, we find that JSNS$^2$ is likely to provide unsurpassed sensitivity, since other facilities suffer from low $K^+$ yields (e.g.\ at a PIP-II beam dump~\cite{Toups:2022yxs}, LANSCE~\cite{CCM:2021yzc}, or the SNS \cite{COHERENT:2019kwz}) or do not have competitive intensity (e.g.\ the SBN beam dump concept~\cite{Toups:2022knq}). Experiments with detectors far downstream also suffer from large $1/L^2$ geometric suppressions {\it c.f.} \cref{a-rate}. Modified experimental designs, e.g.\ a PIP-II beam dump with a proton beam energy $T_p\gtrsim 2~{\rm GeV}$, could allow for competitive KDAR rates. A large volume detector placed near the DUNE hadron absorber or coupled with a high-intensity 8 GeV beam for a muon collider demonstrator would both offer promising future sensitivity. Nevertheless, with JSNS$^2$ already taking data~\cite{Maruyama:2022juu}, there is an immediate opportunity to shed light on heavy axion models in currently unprobed regions of parameter space.

We strongly encourage the JSNS$^2$ collaboration to incorporate axion searches into their central physics program. The signal we have identified, will generically lie far above the signal windows of interest for neutrino physics. Kaon decay at rest produces a monoenergetic $\nu_\mu$ with $E_\nu = 236~{\rm MeV}$, however the resultant muon signature is very different from the signal we have identified heerein. If an axion is produced and decays visibly, the signal is two collimated photons with a total energy between 227~MeV and 354~MeV. Since this lies outside the range of any planned physics goals at JSNS$^2$ this signal could be easily missed; it should not be and a dedicated search should be performed.
For instance, our background estimate relies on additional lead shielding which is absent in reality as we have mentioned.
Therefore, if JSNS$^2$ collaboration observes a larger background than the estimate in~\cite{Jordan:2018gcd},
we believe that our study provides a strong motivation to, \emph{e.g.}, install additional shielding to suppress background 
and to explore the potential signals coming from the axions.

\textbf{Note added:} After the completion of this manuscript we were made aware of other experiments in the J-PARC facility which could search for axions from KDAR \cite{Matheus_Private}. These include KOTO and ND280. It would be interesting to better understand the capability of these experiments to do searches using the JSNS$^2$ beam stop. 

\pagebreak

\textbf{Acknowledgments:} This project arose from discussions at the Aspen Center for Physics (supported by National Science Foundation grant PHY-2210452) and we would like to collectively thank the center for their hospitality and support. We thank Josh Spitz and Takasumi Maruyama for useful correspondence and feedback, and Bertrand Echenard for suggestions involving future muon facilities. YE and ZL are supported in part by the DOE grant DE-SC0011842. RP is supported by the Neutrino Theory Network under Award Number DEAC02-07CHI11359, the U.S. Department of Energy, Office of Science, Office of High Energy Physics, under Award Number DE-SC0011632, and by the Walter Burke Institute for Theoretical Physics.


\bibliography{ref.bib}

\begin{thebibliography}{99}%
\makeatletter
\providecommand \@ifxundefined [1]{%
 \@ifx{#1\undefined}
}%
\providecommand \@ifnum [1]{%
 \ifnum #1\expandafter \@firstoftwo
 \else \expandafter \@secondoftwo
 \fi
}%
\providecommand \@ifx [1]{%
 \ifx #1\expandafter \@firstoftwo
 \else \expandafter \@secondoftwo
 \fi
}%
\providecommand \natexlab [1]{#1}%
\providecommand \enquote  [1]{``#1''}%
\providecommand \bibnamefont  [1]{#1}%
\providecommand \bibfnamefont [1]{#1}%
\providecommand \citenamefont [1]{#1}%
\providecommand \href@noop [0]{\@secondoftwo}%
\providecommand \href [0]{\begingroup \@sanitize@url \@href}%
\providecommand \@href[1]{\@@startlink{#1}\@@href}%
\providecommand \@@href[1]{\endgroup#1\@@endlink}%
\providecommand \@sanitize@url [0]{\catcode `\\12\catcode `\$12\catcode
  `\&12\catcode `\#12\catcode `\^12\catcode `\_12\catcode `\%12\relax}%
\providecommand \@@startlink[1]{}%
\providecommand \@@endlink[0]{}%
\providecommand \url  [0]{\begingroup\@sanitize@url \@url }%
\providecommand \@url [1]{\endgroup\@href {#1}{\urlprefix }}%
\providecommand \urlprefix  [0]{URL }%
\providecommand \Eprint [0]{\href }%
\providecommand \doibase [0]{http://dx.doi.org/}%
\providecommand \selectlanguage [0]{\@gobble}%
\providecommand \bibinfo  [0]{\@secondoftwo}%
\providecommand \bibfield  [0]{\@secondoftwo}%
\providecommand \translation [1]{[#1]}%
\providecommand \BibitemOpen [0]{}%
\providecommand \bibitemStop [0]{}%
\providecommand \bibitemNoStop [0]{.\EOS\space}%
\providecommand \EOS [0]{\spacefactor3000\relax}%
\providecommand \BibitemShut  [1]{\csname bibitem#1\endcsname}%
\let\auto@bib@innerbib\@empty
\bibitem [{\citenamefont {Graner}\ \emph {et~al.}(2016)\citenamefont {Graner},
  \citenamefont {Chen}, \citenamefont {Lindahl},\ and\ \citenamefont
  {Heckel}}]{Graner:2016ses}%
  \BibitemOpen
  \bibfield  {author} {\bibinfo {author} {\bibfnamefont {B.}~\bibnamefont
  {Graner}}, \bibinfo {author} {\bibfnamefont {Y.}~\bibnamefont {Chen}},
  \bibinfo {author} {\bibfnamefont {E.~G.}\ \bibnamefont {Lindahl}}, \ and\
  \bibinfo {author} {\bibfnamefont {B.~R.}\ \bibnamefont {Heckel}},\ }\bibfield
   {title} {\enquote {\bibinfo {title} {{Reduced Limit on the Permanent
  Electric Dipole Moment of Hg199}},}\ }\href {\doibase
  10.1103/PhysRevLett.116.161601} {\bibfield  {journal} {\bibinfo  {journal}
  {Phys. Rev. Lett.}\ }\textbf {\bibinfo {volume} {116}},\ \bibinfo {pages}
  {161601} (\bibinfo {year} {2016})},\ \bibinfo {note} {[Erratum:
  Phys.Rev.Lett. 119, 119901 (2017)]},\ \Eprint
  {http://arxiv.org/abs/1601.04339} {arXiv:1601.04339 [physics.atom-ph]}
  \BibitemShut {NoStop}%
\bibitem [{\citenamefont {Abel}\ \emph {et~al.}(2020)\citenamefont {Abel} \emph
  {et~al.}}]{Abel:2020gbr}%
  \BibitemOpen
  \bibfield  {author} {\bibinfo {author} {\bibfnamefont {C.}~\bibnamefont
  {Abel}} \emph {et~al.},\ }\bibfield  {title} {\enquote {\bibinfo {title}
  {{Measurement of the Permanent Electric Dipole Moment of the Neutron}},}\
  }\href {\doibase 10.1103/PhysRevLett.124.081803} {\bibfield  {journal}
  {\bibinfo  {journal} {Phys. Rev. Lett.}\ }\textbf {\bibinfo {volume} {124}},\
  \bibinfo {pages} {081803} (\bibinfo {year} {2020})},\ \Eprint
  {http://arxiv.org/abs/2001.11966} {arXiv:2001.11966 [hep-ex]} \BibitemShut
  {NoStop}%
\bibitem [{\citenamefont {Peccei}\ and\ \citenamefont
  {Quinn}(1977{\natexlab{a}})}]{Peccei:1977hh}%
  \BibitemOpen
  \bibfield  {author} {\bibinfo {author} {\bibfnamefont {R.~D.}\ \bibnamefont
  {Peccei}}\ and\ \bibinfo {author} {\bibfnamefont {Helen~R.}\ \bibnamefont
  {Quinn}},\ }\bibfield  {title} {\enquote {\bibinfo {title} {{CP Conservation
  in the Presence of Instantons}},}\ }\href {\doibase
  10.1103/PhysRevLett.38.1440} {\bibfield  {journal} {\bibinfo  {journal}
  {Phys. Rev. Lett.}\ }\textbf {\bibinfo {volume} {38}},\ \bibinfo {pages}
  {1440--1443} (\bibinfo {year} {1977}{\natexlab{a}})}\BibitemShut {NoStop}%
\bibitem [{\citenamefont {Peccei}\ and\ \citenamefont
  {Quinn}(1977{\natexlab{b}})}]{Peccei:1977ur}%
  \BibitemOpen
  \bibfield  {author} {\bibinfo {author} {\bibfnamefont {R.~D.}\ \bibnamefont
  {Peccei}}\ and\ \bibinfo {author} {\bibfnamefont {Helen~R.}\ \bibnamefont
  {Quinn}},\ }\bibfield  {title} {\enquote {\bibinfo {title} {{Constraints
  Imposed by CP Conservation in the Presence of Instantons}},}\ }\href
  {\doibase 10.1103/PhysRevD.16.1791} {\bibfield  {journal} {\bibinfo
  {journal} {Phys. Rev. D}\ }\textbf {\bibinfo {volume} {16}},\ \bibinfo
  {pages} {1791--1797} (\bibinfo {year} {1977}{\natexlab{b}})}\BibitemShut
  {NoStop}%
\bibitem [{\citenamefont {Weinberg}(1978)}]{Weinberg:1977ma}%
  \BibitemOpen
  \bibfield  {author} {\bibinfo {author} {\bibfnamefont {Steven}\ \bibnamefont
  {Weinberg}},\ }\bibfield  {title} {\enquote {\bibinfo {title} {{A New Light
  Boson?}}}\ }\href {\doibase 10.1103/PhysRevLett.40.223} {\bibfield  {journal}
  {\bibinfo  {journal} {Phys. Rev. Lett.}\ }\textbf {\bibinfo {volume} {40}},\
  \bibinfo {pages} {223--226} (\bibinfo {year} {1978})}\BibitemShut {NoStop}%
\bibitem [{\citenamefont {Wilczek}(1978)}]{Wilczek:1977pj}%
  \BibitemOpen
  \bibfield  {author} {\bibinfo {author} {\bibfnamefont {Frank}\ \bibnamefont
  {Wilczek}},\ }\bibfield  {title} {\enquote {\bibinfo {title} {{Problem of
  Strong $P$ and $T$ Invariance in the Presence of Instantons}},}\ }\href
  {\doibase 10.1103/PhysRevLett.40.279} {\bibfield  {journal} {\bibinfo
  {journal} {Phys. Rev. Lett.}\ }\textbf {\bibinfo {volume} {40}},\ \bibinfo
  {pages} {279--282} (\bibinfo {year} {1978})}\BibitemShut {NoStop}%
\bibitem [{\citenamefont {Kamionkowski}\ and\ \citenamefont
  {March-Russell}(1992)}]{Kamionkowski:1992mf}%
  \BibitemOpen
  \bibfield  {author} {\bibinfo {author} {\bibfnamefont {Marc}\ \bibnamefont
  {Kamionkowski}}\ and\ \bibinfo {author} {\bibfnamefont {John}\ \bibnamefont
  {March-Russell}},\ }\bibfield  {title} {\enquote {\bibinfo {title} {{Planck
  scale physics and the Peccei-Quinn mechanism}},}\ }\href {\doibase
  10.1016/0370-2693(92)90492-M} {\bibfield  {journal} {\bibinfo  {journal}
  {Phys. Lett. B}\ }\textbf {\bibinfo {volume} {282}},\ \bibinfo {pages}
  {137--141} (\bibinfo {year} {1992})},\ \Eprint
  {http://arxiv.org/abs/hep-th/9202003} {arXiv:hep-th/9202003} \BibitemShut
  {NoStop}%
\bibitem [{\citenamefont {Barr}\ and\ \citenamefont
  {Seckel}(1992)}]{Barr:1992qq}%
  \BibitemOpen
  \bibfield  {author} {\bibinfo {author} {\bibfnamefont {Stephen~M.}\
  \bibnamefont {Barr}}\ and\ \bibinfo {author} {\bibfnamefont {D.}~\bibnamefont
  {Seckel}},\ }\bibfield  {title} {\enquote {\bibinfo {title} {{Planck scale
  corrections to axion models}},}\ }\href {\doibase 10.1103/PhysRevD.46.539}
  {\bibfield  {journal} {\bibinfo  {journal} {Phys. Rev. D}\ }\textbf {\bibinfo
  {volume} {46}},\ \bibinfo {pages} {539--549} (\bibinfo {year}
  {1992})}\BibitemShut {NoStop}%
\bibitem [{\citenamefont {Ghigna}\ \emph {et~al.}(1992)\citenamefont {Ghigna},
  \citenamefont {Lusignoli},\ and\ \citenamefont {Roncadelli}}]{Ghigna:1992iv}%
  \BibitemOpen
  \bibfield  {author} {\bibinfo {author} {\bibfnamefont {S.}~\bibnamefont
  {Ghigna}}, \bibinfo {author} {\bibfnamefont {Maurizio}\ \bibnamefont
  {Lusignoli}}, \ and\ \bibinfo {author} {\bibfnamefont {M.}~\bibnamefont
  {Roncadelli}},\ }\bibfield  {title} {\enquote {\bibinfo {title} {{Instability
  of the invisible axion}},}\ }\href {\doibase 10.1016/0370-2693(92)90019-Z}
  {\bibfield  {journal} {\bibinfo  {journal} {Phys. Lett. B}\ }\textbf
  {\bibinfo {volume} {283}},\ \bibinfo {pages} {278--281} (\bibinfo {year}
  {1992})}\BibitemShut {NoStop}%
\bibitem [{\citenamefont {Holman}\ \emph {et~al.}(1992)\citenamefont {Holman},
  \citenamefont {Hsu}, \citenamefont {Kephart}, \citenamefont {Kolb},
  \citenamefont {Watkins},\ and\ \citenamefont {Widrow}}]{Holman:1992us}%
  \BibitemOpen
  \bibfield  {author} {\bibinfo {author} {\bibfnamefont {Richard}\ \bibnamefont
  {Holman}}, \bibinfo {author} {\bibfnamefont {Stephen D.~H.}\ \bibnamefont
  {Hsu}}, \bibinfo {author} {\bibfnamefont {Thomas~W.}\ \bibnamefont
  {Kephart}}, \bibinfo {author} {\bibfnamefont {Edward~W.}\ \bibnamefont
  {Kolb}}, \bibinfo {author} {\bibfnamefont {Richard}\ \bibnamefont {Watkins}},
  \ and\ \bibinfo {author} {\bibfnamefont {Lawrence~M.}\ \bibnamefont
  {Widrow}},\ }\bibfield  {title} {\enquote {\bibinfo {title} {{Solutions to
  the strong CP problem in a world with gravity}},}\ }\href {\doibase
  10.1016/0370-2693(92)90491-L} {\bibfield  {journal} {\bibinfo  {journal}
  {Phys. Lett. B}\ }\textbf {\bibinfo {volume} {282}},\ \bibinfo {pages}
  {132--136} (\bibinfo {year} {1992})},\ \Eprint
  {http://arxiv.org/abs/hep-ph/9203206} {arXiv:hep-ph/9203206} \BibitemShut
  {NoStop}%
\bibitem [{\citenamefont {Dimopoulos}(1979)}]{Dimopoulos:1979pp}%
  \BibitemOpen
  \bibfield  {author} {\bibinfo {author} {\bibfnamefont {Savas}\ \bibnamefont
  {Dimopoulos}},\ }\bibfield  {title} {\enquote {\bibinfo {title} {{A Solution
  of the Strong {CP} Problem in Models With Scalars}},}\ }\href {\doibase
  10.1016/0370-2693(79)91233-4} {\bibfield  {journal} {\bibinfo  {journal}
  {Phys. Lett. B}\ }\textbf {\bibinfo {volume} {84}},\ \bibinfo {pages}
  {435--439} (\bibinfo {year} {1979})}\BibitemShut {NoStop}%
\bibitem [{\citenamefont {Tye}(1981)}]{Tye:1981zy}%
  \BibitemOpen
  \bibfield  {author} {\bibinfo {author} {\bibfnamefont {S.~H.~H.}\
  \bibnamefont {Tye}},\ }\bibfield  {title} {\enquote {\bibinfo {title} {{A
  Superstrong Force With a Heavy Axion}},}\ }\href {\doibase
  10.1103/PhysRevLett.47.1035} {\bibfield  {journal} {\bibinfo  {journal}
  {Phys. Rev. Lett.}\ }\textbf {\bibinfo {volume} {47}},\ \bibinfo {pages}
  {1035} (\bibinfo {year} {1981})}\BibitemShut {NoStop}%
\bibitem [{\citenamefont {Holdom}\ and\ \citenamefont
  {Peskin}(1982)}]{Holdom:1982ex}%
  \BibitemOpen
  \bibfield  {author} {\bibinfo {author} {\bibfnamefont {Bob}\ \bibnamefont
  {Holdom}}\ and\ \bibinfo {author} {\bibfnamefont {Michael~E.}\ \bibnamefont
  {Peskin}},\ }\bibfield  {title} {\enquote {\bibinfo {title} {{Raising the
  Axion Mass}},}\ }\href {\doibase 10.1016/0550-3213(82)90228-0} {\bibfield
  {journal} {\bibinfo  {journal} {Nucl. Phys. B}\ }\textbf {\bibinfo {volume}
  {208}},\ \bibinfo {pages} {397--412} (\bibinfo {year} {1982})}\BibitemShut
  {NoStop}%
\bibitem [{\citenamefont {Flynn}\ and\ \citenamefont
  {Randall}(1987)}]{Flynn:1987rs}%
  \BibitemOpen
  \bibfield  {author} {\bibinfo {author} {\bibfnamefont {Jonathan~M.}\
  \bibnamefont {Flynn}}\ and\ \bibinfo {author} {\bibfnamefont {Lisa}\
  \bibnamefont {Randall}},\ }\bibfield  {title} {\enquote {\bibinfo {title} {{A
  Computation of the Small Instanton Contribution to the Axion Potential}},}\
  }\href {\doibase 10.1016/0550-3213(87)90089-7} {\bibfield  {journal}
  {\bibinfo  {journal} {Nucl. Phys. B}\ }\textbf {\bibinfo {volume} {293}},\
  \bibinfo {pages} {731--739} (\bibinfo {year} {1987})}\BibitemShut {NoStop}%
\bibitem [{\citenamefont {Rubakov}(1997)}]{Rubakov:1997vp}%
  \BibitemOpen
  \bibfield  {author} {\bibinfo {author} {\bibfnamefont {V.~A.}\ \bibnamefont
  {Rubakov}},\ }\bibfield  {title} {\enquote {\bibinfo {title} {{Grand
  unification and heavy axion}},}\ }\href {\doibase 10.1134/1.567390}
  {\bibfield  {journal} {\bibinfo  {journal} {JETP Lett.}\ }\textbf {\bibinfo
  {volume} {65}},\ \bibinfo {pages} {621--624} (\bibinfo {year} {1997})},\
  \Eprint {http://arxiv.org/abs/hep-ph/9703409} {arXiv:hep-ph/9703409}
  \BibitemShut {NoStop}%
\bibitem [{\citenamefont {Berezhiani}\ \emph {et~al.}(2001)\citenamefont
  {Berezhiani}, \citenamefont {Gianfagna},\ and\ \citenamefont
  {Giannotti}}]{Berezhiani:2000gh}%
  \BibitemOpen
  \bibfield  {author} {\bibinfo {author} {\bibfnamefont {Zurab}\ \bibnamefont
  {Berezhiani}}, \bibinfo {author} {\bibfnamefont {Leonida}\ \bibnamefont
  {Gianfagna}}, \ and\ \bibinfo {author} {\bibfnamefont {Maurizio}\
  \bibnamefont {Giannotti}},\ }\bibfield  {title} {\enquote {\bibinfo {title}
  {{Strong CP problem and mirror world: The Weinberg-Wilczek axion
  revisited}},}\ }\href {\doibase 10.1016/S0370-2693(00)01392-7} {\bibfield
  {journal} {\bibinfo  {journal} {Phys. Lett. B}\ }\textbf {\bibinfo {volume}
  {500}},\ \bibinfo {pages} {286--296} (\bibinfo {year} {2001})},\ \Eprint
  {http://arxiv.org/abs/hep-ph/0009290} {arXiv:hep-ph/0009290} \BibitemShut
  {NoStop}%
\bibitem [{\citenamefont {Hook}(2015)}]{Hook:2014cda}%
  \BibitemOpen
  \bibfield  {author} {\bibinfo {author} {\bibfnamefont {Anson}\ \bibnamefont
  {Hook}},\ }\bibfield  {title} {\enquote {\bibinfo {title} {{Anomalous
  solutions to the strong CP problem}},}\ }\href {\doibase
  10.1103/PhysRevLett.114.141801} {\bibfield  {journal} {\bibinfo  {journal}
  {Phys. Rev. Lett.}\ }\textbf {\bibinfo {volume} {114}},\ \bibinfo {pages}
  {141801} (\bibinfo {year} {2015})},\ \Eprint {http://arxiv.org/abs/1411.3325}
  {arXiv:1411.3325 [hep-ph]} \BibitemShut {NoStop}%
\bibitem [{\citenamefont {Fukuda}\ \emph {et~al.}(2015)\citenamefont {Fukuda},
  \citenamefont {Harigaya}, \citenamefont {Ibe},\ and\ \citenamefont
  {Yanagida}}]{Fukuda:2015ana}%
  \BibitemOpen
  \bibfield  {author} {\bibinfo {author} {\bibfnamefont {Hajime}\ \bibnamefont
  {Fukuda}}, \bibinfo {author} {\bibfnamefont {Keisuke}\ \bibnamefont
  {Harigaya}}, \bibinfo {author} {\bibfnamefont {Masahiro}\ \bibnamefont
  {Ibe}}, \ and\ \bibinfo {author} {\bibfnamefont {Tsutomu~T.}\ \bibnamefont
  {Yanagida}},\ }\bibfield  {title} {\enquote {\bibinfo {title} {{Model of
  visible QCD axion}},}\ }\href {\doibase 10.1103/PhysRevD.92.015021}
  {\bibfield  {journal} {\bibinfo  {journal} {Phys. Rev. D}\ }\textbf {\bibinfo
  {volume} {92}},\ \bibinfo {pages} {015021} (\bibinfo {year} {2015})},\
  \Eprint {http://arxiv.org/abs/1504.06084} {arXiv:1504.06084 [hep-ph]}
  \BibitemShut {NoStop}%
\bibitem [{\citenamefont {Gherghetta}\ \emph {et~al.}(2016)\citenamefont
  {Gherghetta}, \citenamefont {Nagata},\ and\ \citenamefont
  {Shifman}}]{Gherghetta:2016fhp}%
  \BibitemOpen
  \bibfield  {author} {\bibinfo {author} {\bibfnamefont {Tony}\ \bibnamefont
  {Gherghetta}}, \bibinfo {author} {\bibfnamefont {Natsumi}\ \bibnamefont
  {Nagata}}, \ and\ \bibinfo {author} {\bibfnamefont {Mikhail}\ \bibnamefont
  {Shifman}},\ }\bibfield  {title} {\enquote {\bibinfo {title} {{A Visible QCD
  Axion from an Enlarged Color Group}},}\ }\href {\doibase
  10.1103/PhysRevD.93.115010} {\bibfield  {journal} {\bibinfo  {journal} {Phys.
  Rev. D}\ }\textbf {\bibinfo {volume} {93}},\ \bibinfo {pages} {115010}
  (\bibinfo {year} {2016})},\ \Eprint {http://arxiv.org/abs/1604.01127}
  {arXiv:1604.01127 [hep-ph]} \BibitemShut {NoStop}%
\bibitem [{\citenamefont {Dimopoulos}\ \emph {et~al.}(2016)\citenamefont
  {Dimopoulos}, \citenamefont {Hook}, \citenamefont {Huang},\ and\
  \citenamefont {Marques-Tavares}}]{Dimopoulos:2016lvn}%
  \BibitemOpen
  \bibfield  {author} {\bibinfo {author} {\bibfnamefont {Savas}\ \bibnamefont
  {Dimopoulos}}, \bibinfo {author} {\bibfnamefont {Anson}\ \bibnamefont
  {Hook}}, \bibinfo {author} {\bibfnamefont {Junwu}\ \bibnamefont {Huang}}, \
  and\ \bibinfo {author} {\bibfnamefont {Gustavo}\ \bibnamefont
  {Marques-Tavares}},\ }\bibfield  {title} {\enquote {\bibinfo {title} {{A
  collider observable QCD axion}},}\ }\href {\doibase 10.1007/JHEP11(2016)052}
  {\bibfield  {journal} {\bibinfo  {journal} {JHEP}\ }\textbf {\bibinfo
  {volume} {11}},\ \bibinfo {pages} {052} (\bibinfo {year} {2016})},\ \Eprint
  {http://arxiv.org/abs/1606.03097} {arXiv:1606.03097 [hep-ph]} \BibitemShut
  {NoStop}%
\bibitem [{\citenamefont {Agrawal}\ and\ \citenamefont
  {Howe}(2018{\natexlab{a}})}]{Agrawal:2017ksf}%
  \BibitemOpen
  \bibfield  {author} {\bibinfo {author} {\bibfnamefont {Prateek}\ \bibnamefont
  {Agrawal}}\ and\ \bibinfo {author} {\bibfnamefont {Kiel}\ \bibnamefont
  {Howe}},\ }\bibfield  {title} {\enquote {\bibinfo {title} {{Factoring the
  Strong CP Problem}},}\ }\href {\doibase 10.1007/JHEP12(2018)029} {\bibfield
  {journal} {\bibinfo  {journal} {JHEP}\ }\textbf {\bibinfo {volume} {12}},\
  \bibinfo {pages} {029} (\bibinfo {year} {2018}{\natexlab{a}})},\ \Eprint
  {http://arxiv.org/abs/1710.04213} {arXiv:1710.04213 [hep-ph]} \BibitemShut
  {NoStop}%
\bibitem [{\citenamefont {Agrawal}\ and\ \citenamefont
  {Howe}(2018{\natexlab{b}})}]{Agrawal:2017evu}%
  \BibitemOpen
  \bibfield  {author} {\bibinfo {author} {\bibfnamefont {Prateek}\ \bibnamefont
  {Agrawal}}\ and\ \bibinfo {author} {\bibfnamefont {Kiel}\ \bibnamefont
  {Howe}},\ }\bibfield  {title} {\enquote {\bibinfo {title} {{A Flavorful
  Factoring of the Strong CP Problem}},}\ }\href {\doibase
  10.1007/JHEP12(2018)035} {\bibfield  {journal} {\bibinfo  {journal} {JHEP}\
  }\textbf {\bibinfo {volume} {12}},\ \bibinfo {pages} {035} (\bibinfo {year}
  {2018}{\natexlab{b}})},\ \Eprint {http://arxiv.org/abs/1712.05803}
  {arXiv:1712.05803 [hep-ph]} \BibitemShut {NoStop}%
\bibitem [{\citenamefont {Gaillard}\ \emph {et~al.}(2018)\citenamefont
  {Gaillard}, \citenamefont {Gavela}, \citenamefont {Houtz}, \citenamefont
  {Quilez},\ and\ \citenamefont {Del~Rey}}]{Gaillard:2018xgk}%
  \BibitemOpen
  \bibfield  {author} {\bibinfo {author} {\bibfnamefont {M.~K.}\ \bibnamefont
  {Gaillard}}, \bibinfo {author} {\bibfnamefont {M.~B.}\ \bibnamefont
  {Gavela}}, \bibinfo {author} {\bibfnamefont {R.}~\bibnamefont {Houtz}},
  \bibinfo {author} {\bibfnamefont {P.}~\bibnamefont {Quilez}}, \ and\ \bibinfo
  {author} {\bibfnamefont {R.}~\bibnamefont {Del~Rey}},\ }\bibfield  {title}
  {\enquote {\bibinfo {title} {{Color unified dynamical axion}},}\ }\href
  {\doibase 10.1140/epjc/s10052-018-6396-6} {\bibfield  {journal} {\bibinfo
  {journal} {Eur. Phys. J. C}\ }\textbf {\bibinfo {volume} {78}},\ \bibinfo
  {pages} {972} (\bibinfo {year} {2018})},\ \Eprint
  {http://arxiv.org/abs/1805.06465} {arXiv:1805.06465 [hep-ph]} \BibitemShut
  {NoStop}%
\bibitem [{\citenamefont {Lillard}\ and\ \citenamefont
  {Tait}(2018)}]{Lillard:2018fdt}%
  \BibitemOpen
  \bibfield  {author} {\bibinfo {author} {\bibfnamefont {Benjamin}\
  \bibnamefont {Lillard}}\ and\ \bibinfo {author} {\bibfnamefont {Tim M.~P.}\
  \bibnamefont {Tait}},\ }\bibfield  {title} {\enquote {\bibinfo {title} {{A
  High Quality Composite Axion}},}\ }\href {\doibase 10.1007/JHEP11(2018)199}
  {\bibfield  {journal} {\bibinfo  {journal} {JHEP}\ }\textbf {\bibinfo
  {volume} {11}},\ \bibinfo {pages} {199} (\bibinfo {year} {2018})},\ \Eprint
  {http://arxiv.org/abs/1811.03089} {arXiv:1811.03089 [hep-ph]} \BibitemShut
  {NoStop}%
\bibitem [{\citenamefont {Fuentes-Mart\'\i{}n}\ \emph
  {et~al.}(2019)\citenamefont {Fuentes-Mart\'\i{}n}, \citenamefont {Reig},\
  and\ \citenamefont {Vicente}}]{Fuentes-Martin:2019bue}%
  \BibitemOpen
  \bibfield  {author} {\bibinfo {author} {\bibfnamefont {Javier}\ \bibnamefont
  {Fuentes-Mart\'\i{}n}}, \bibinfo {author} {\bibfnamefont {Mario}\
  \bibnamefont {Reig}}, \ and\ \bibinfo {author} {\bibfnamefont {Avelino}\
  \bibnamefont {Vicente}},\ }\bibfield  {title} {\enquote {\bibinfo {title}
  {{Strong $CP$ problem with low-energy emergent QCD: The 4321 case}},}\ }\href
  {\doibase 10.1103/PhysRevD.100.115028} {\bibfield  {journal} {\bibinfo
  {journal} {Phys. Rev. D}\ }\textbf {\bibinfo {volume} {100}},\ \bibinfo
  {pages} {115028} (\bibinfo {year} {2019})},\ \Eprint
  {http://arxiv.org/abs/1907.02550} {arXiv:1907.02550 [hep-ph]} \BibitemShut
  {NoStop}%
\bibitem [{\citenamefont {Cs\'aki}\ \emph {et~al.}(2020)\citenamefont
  {Cs\'aki}, \citenamefont {Ruhdorfer},\ and\ \citenamefont
  {Shirman}}]{Csaki:2019vte}%
  \BibitemOpen
  \bibfield  {author} {\bibinfo {author} {\bibfnamefont {Csaba}\ \bibnamefont
  {Cs\'aki}}, \bibinfo {author} {\bibfnamefont {Maximilian}\ \bibnamefont
  {Ruhdorfer}}, \ and\ \bibinfo {author} {\bibfnamefont {Yuri}\ \bibnamefont
  {Shirman}},\ }\bibfield  {title} {\enquote {\bibinfo {title} {{UV Sensitivity
  of the Axion Mass from Instantons in Partially Broken Gauge Groups}},}\
  }\href {\doibase 10.1007/JHEP04(2020)031} {\bibfield  {journal} {\bibinfo
  {journal} {JHEP}\ }\textbf {\bibinfo {volume} {04}},\ \bibinfo {pages} {031}
  (\bibinfo {year} {2020})},\ \Eprint {http://arxiv.org/abs/1912.02197}
  {arXiv:1912.02197 [hep-ph]} \BibitemShut {NoStop}%
\bibitem [{\citenamefont {Hook}\ \emph {et~al.}(2020)\citenamefont {Hook},
  \citenamefont {Kumar}, \citenamefont {Liu},\ and\ \citenamefont
  {Sundrum}}]{Hook:2019qoh}%
  \BibitemOpen
  \bibfield  {author} {\bibinfo {author} {\bibfnamefont {Anson}\ \bibnamefont
  {Hook}}, \bibinfo {author} {\bibfnamefont {Soubhik}\ \bibnamefont {Kumar}},
  \bibinfo {author} {\bibfnamefont {Zhen}\ \bibnamefont {Liu}}, \ and\ \bibinfo
  {author} {\bibfnamefont {Raman}\ \bibnamefont {Sundrum}},\ }\bibfield
  {title} {\enquote {\bibinfo {title} {{High Quality QCD Axion and the LHC}},}\
  }\href {\doibase 10.1103/PhysRevLett.124.221801} {\bibfield  {journal}
  {\bibinfo  {journal} {Phys. Rev. Lett.}\ }\textbf {\bibinfo {volume} {124}},\
  \bibinfo {pages} {221801} (\bibinfo {year} {2020})},\ \Eprint
  {http://arxiv.org/abs/1911.12364} {arXiv:1911.12364 [hep-ph]} \BibitemShut
  {NoStop}%
\bibitem [{\citenamefont {Gherghetta}\ \emph {et~al.}(2020)\citenamefont
  {Gherghetta}, \citenamefont {Khoze}, \citenamefont {Pomarol},\ and\
  \citenamefont {Shirman}}]{Gherghetta:2020keg}%
  \BibitemOpen
  \bibfield  {author} {\bibinfo {author} {\bibfnamefont {Tony}\ \bibnamefont
  {Gherghetta}}, \bibinfo {author} {\bibfnamefont {Valentin~V.}\ \bibnamefont
  {Khoze}}, \bibinfo {author} {\bibfnamefont {Alex}\ \bibnamefont {Pomarol}}, \
  and\ \bibinfo {author} {\bibfnamefont {Yuri}\ \bibnamefont {Shirman}},\
  }\bibfield  {title} {\enquote {\bibinfo {title} {{The Axion Mass from 5D
  Small Instantons}},}\ }\href {\doibase 10.1007/JHEP03(2020)063} {\bibfield
  {journal} {\bibinfo  {journal} {JHEP}\ }\textbf {\bibinfo {volume} {03}},\
  \bibinfo {pages} {063} (\bibinfo {year} {2020})},\ \Eprint
  {http://arxiv.org/abs/2001.05610} {arXiv:2001.05610 [hep-ph]} \BibitemShut
  {NoStop}%
\bibitem [{\citenamefont {Gherghetta}\ and\ \citenamefont
  {Nguyen}(2020)}]{Gherghetta:2020ofz}%
  \BibitemOpen
  \bibfield  {author} {\bibinfo {author} {\bibfnamefont {Tony}\ \bibnamefont
  {Gherghetta}}\ and\ \bibinfo {author} {\bibfnamefont {Minh~D.}\ \bibnamefont
  {Nguyen}},\ }\bibfield  {title} {\enquote {\bibinfo {title} {{A Composite
  Higgs with a Heavy Composite Axion}},}\ }\href {\doibase
  10.1007/JHEP12(2020)094} {\bibfield  {journal} {\bibinfo  {journal} {JHEP}\
  }\textbf {\bibinfo {volume} {12}},\ \bibinfo {pages} {094} (\bibinfo {year}
  {2020})},\ \Eprint {http://arxiv.org/abs/2007.10875} {arXiv:2007.10875
  [hep-ph]} \BibitemShut {NoStop}%
\bibitem [{\citenamefont {Valenti}\ \emph {et~al.}(2022)\citenamefont
  {Valenti}, \citenamefont {Vecchi},\ and\ \citenamefont
  {Xu}}]{Valenti:2022tsc}%
  \BibitemOpen
  \bibfield  {author} {\bibinfo {author} {\bibfnamefont {Alessandro}\
  \bibnamefont {Valenti}}, \bibinfo {author} {\bibfnamefont {Luca}\
  \bibnamefont {Vecchi}}, \ and\ \bibinfo {author} {\bibfnamefont {Ling-Xiao}\
  \bibnamefont {Xu}},\ }\bibfield  {title} {\enquote {\bibinfo {title} {{Grand
  Color axion}},}\ }\href {\doibase 10.1007/JHEP10(2022)025} {\bibfield
  {journal} {\bibinfo  {journal} {JHEP}\ }\textbf {\bibinfo {volume} {10}},\
  \bibinfo {pages} {025} (\bibinfo {year} {2022})},\ \Eprint
  {http://arxiv.org/abs/2206.04077} {arXiv:2206.04077 [hep-ph]} \BibitemShut
  {NoStop}%
\bibitem [{\citenamefont {Kivel}\ \emph {et~al.}(2022)\citenamefont {Kivel},
  \citenamefont {Laux},\ and\ \citenamefont {Yu}}]{Kivel:2022emq}%
  \BibitemOpen
  \bibfield  {author} {\bibinfo {author} {\bibfnamefont {Alexey}\ \bibnamefont
  {Kivel}}, \bibinfo {author} {\bibfnamefont {Julien}\ \bibnamefont {Laux}}, \
  and\ \bibinfo {author} {\bibfnamefont {Felix}\ \bibnamefont {Yu}},\
  }\bibfield  {title} {\enquote {\bibinfo {title} {{Supersizing axions with
  small size instantons}},}\ }\href {\doibase 10.1007/JHEP11(2022)088}
  {\bibfield  {journal} {\bibinfo  {journal} {JHEP}\ }\textbf {\bibinfo
  {volume} {11}},\ \bibinfo {pages} {088} (\bibinfo {year} {2022})},\ \Eprint
  {http://arxiv.org/abs/2207.08740} {arXiv:2207.08740 [hep-ph]} \BibitemShut
  {NoStop}%
\bibitem [{\citenamefont {Bauer}\ \emph {et~al.}(2017)\citenamefont {Bauer},
  \citenamefont {Neubert},\ and\ \citenamefont {Thamm}}]{Bauer:2017ris}%
  \BibitemOpen
  \bibfield  {author} {\bibinfo {author} {\bibfnamefont {Martin}\ \bibnamefont
  {Bauer}}, \bibinfo {author} {\bibfnamefont {Matthias}\ \bibnamefont
  {Neubert}}, \ and\ \bibinfo {author} {\bibfnamefont {Andrea}\ \bibnamefont
  {Thamm}},\ }\bibfield  {title} {\enquote {\bibinfo {title} {{Collider Probes
  of Axion-Like Particles}},}\ }\href {\doibase 10.1007/JHEP12(2017)044}
  {\bibfield  {journal} {\bibinfo  {journal} {JHEP}\ }\textbf {\bibinfo
  {volume} {12}},\ \bibinfo {pages} {044} (\bibinfo {year} {2017})},\ \Eprint
  {http://arxiv.org/abs/1708.00443} {arXiv:1708.00443 [hep-ph]} \BibitemShut
  {NoStop}%
\bibitem [{\citenamefont {Irastorza}\ and\ \citenamefont
  {Redondo}(2018)}]{Irastorza:2018dyq}%
  \BibitemOpen
  \bibfield  {author} {\bibinfo {author} {\bibfnamefont {Igor~G.}\ \bibnamefont
  {Irastorza}}\ and\ \bibinfo {author} {\bibfnamefont {Javier}\ \bibnamefont
  {Redondo}},\ }\bibfield  {title} {\enquote {\bibinfo {title} {{New
  experimental approaches in the search for axion-like particles}},}\ }\href
  {\doibase 10.1016/j.ppnp.2018.05.003} {\bibfield  {journal} {\bibinfo
  {journal} {Prog. Part. Nucl. Phys.}\ }\textbf {\bibinfo {volume} {102}},\
  \bibinfo {pages} {89--159} (\bibinfo {year} {2018})},\ \Eprint
  {http://arxiv.org/abs/1801.08127} {arXiv:1801.08127 [hep-ph]} \BibitemShut
  {NoStop}%
\bibitem [{\citenamefont {Arvanitaki}\ \emph {et~al.}(2010)\citenamefont
  {Arvanitaki}, \citenamefont {Dimopoulos}, \citenamefont {Dubovsky},
  \citenamefont {Kaloper},\ and\ \citenamefont
  {March-Russell}}]{Arvanitaki:2009fg}%
  \BibitemOpen
  \bibfield  {author} {\bibinfo {author} {\bibfnamefont {Asimina}\ \bibnamefont
  {Arvanitaki}}, \bibinfo {author} {\bibfnamefont {Savas}\ \bibnamefont
  {Dimopoulos}}, \bibinfo {author} {\bibfnamefont {Sergei}\ \bibnamefont
  {Dubovsky}}, \bibinfo {author} {\bibfnamefont {Nemanja}\ \bibnamefont
  {Kaloper}}, \ and\ \bibinfo {author} {\bibfnamefont {John}\ \bibnamefont
  {March-Russell}},\ }\bibfield  {title} {\enquote {\bibinfo {title} {{String
  Axiverse}},}\ }\href {\doibase 10.1103/PhysRevD.81.123530} {\bibfield
  {journal} {\bibinfo  {journal} {Phys. Rev. D}\ }\textbf {\bibinfo {volume}
  {81}},\ \bibinfo {pages} {123530} (\bibinfo {year} {2010})},\ \Eprint
  {http://arxiv.org/abs/0905.4720} {arXiv:0905.4720 [hep-th]} \BibitemShut
  {NoStop}%
\bibitem [{\citenamefont {Jaeckel}\ and\ \citenamefont
  {Ringwald}(2010)}]{Jaeckel:2010ni}%
  \BibitemOpen
  \bibfield  {author} {\bibinfo {author} {\bibfnamefont {Joerg}\ \bibnamefont
  {Jaeckel}}\ and\ \bibinfo {author} {\bibfnamefont {Andreas}\ \bibnamefont
  {Ringwald}},\ }\bibfield  {title} {\enquote {\bibinfo {title} {{The
  Low-Energy Frontier of Particle Physics}},}\ }\href {\doibase
  10.1146/annurev.nucl.012809.104433} {\bibfield  {journal} {\bibinfo
  {journal} {Ann. Rev. Nucl. Part. Sci.}\ }\textbf {\bibinfo {volume} {60}},\
  \bibinfo {pages} {405--437} (\bibinfo {year} {2010})},\ \Eprint
  {http://arxiv.org/abs/1002.0329} {arXiv:1002.0329 [hep-ph]} \BibitemShut
  {NoStop}%
\bibitem [{\citenamefont {Essig}\ \emph {et~al.}(2010)\citenamefont {Essig},
  \citenamefont {Harnik}, \citenamefont {Kaplan},\ and\ \citenamefont
  {Toro}}]{Essig:2010gu}%
  \BibitemOpen
  \bibfield  {author} {\bibinfo {author} {\bibfnamefont {Rouven}\ \bibnamefont
  {Essig}}, \bibinfo {author} {\bibfnamefont {Roni}\ \bibnamefont {Harnik}},
  \bibinfo {author} {\bibfnamefont {Jared}\ \bibnamefont {Kaplan}}, \ and\
  \bibinfo {author} {\bibfnamefont {Natalia}\ \bibnamefont {Toro}},\ }\bibfield
   {title} {\enquote {\bibinfo {title} {{Discovering New Light States at
  Neutrino Experiments}},}\ }\href {\doibase 10.1103/PhysRevD.82.113008}
  {\bibfield  {journal} {\bibinfo  {journal} {Phys. Rev. D}\ }\textbf {\bibinfo
  {volume} {82}},\ \bibinfo {pages} {113008} (\bibinfo {year} {2010})},\
  \Eprint {http://arxiv.org/abs/1008.0636} {arXiv:1008.0636 [hep-ph]}
  \BibitemShut {NoStop}%
\bibitem [{\citenamefont {D\"obrich}\ \emph {et~al.}(2016)\citenamefont
  {D\"obrich}, \citenamefont {Jaeckel}, \citenamefont {Kahlhoefer},
  \citenamefont {Ringwald},\ and\ \citenamefont
  {Schmidt-Hoberg}}]{Dobrich:2015jyk}%
  \BibitemOpen
  \bibfield  {author} {\bibinfo {author} {\bibfnamefont {Babette}\ \bibnamefont
  {D\"obrich}}, \bibinfo {author} {\bibfnamefont {Joerg}\ \bibnamefont
  {Jaeckel}}, \bibinfo {author} {\bibfnamefont {Felix}\ \bibnamefont
  {Kahlhoefer}}, \bibinfo {author} {\bibfnamefont {Andreas}\ \bibnamefont
  {Ringwald}}, \ and\ \bibinfo {author} {\bibfnamefont {Kai}\ \bibnamefont
  {Schmidt-Hoberg}},\ }\bibfield  {title} {\enquote {\bibinfo {title}
  {{ALPtraum: ALP production in proton beam dump experiments}},}\ }\href
  {\doibase 10.1007/JHEP02(2016)018} {\bibfield  {journal} {\bibinfo  {journal}
  {JHEP}\ }\textbf {\bibinfo {volume} {02}},\ \bibinfo {pages} {018} (\bibinfo
  {year} {2016})},\ \Eprint {http://arxiv.org/abs/1512.03069} {arXiv:1512.03069
  [hep-ph]} \BibitemShut {NoStop}%
\bibitem [{\citenamefont {Dolan}\ \emph {et~al.}(2017)\citenamefont {Dolan},
  \citenamefont {Ferber}, \citenamefont {Hearty}, \citenamefont {Kahlhoefer},\
  and\ \citenamefont {Schmidt-Hoberg}}]{Dolan:2017osp}%
  \BibitemOpen
  \bibfield  {author} {\bibinfo {author} {\bibfnamefont {Matthew~J.}\
  \bibnamefont {Dolan}}, \bibinfo {author} {\bibfnamefont {Torben}\
  \bibnamefont {Ferber}}, \bibinfo {author} {\bibfnamefont {Christopher}\
  \bibnamefont {Hearty}}, \bibinfo {author} {\bibfnamefont {Felix}\
  \bibnamefont {Kahlhoefer}}, \ and\ \bibinfo {author} {\bibfnamefont {Kai}\
  \bibnamefont {Schmidt-Hoberg}},\ }\bibfield  {title} {\enquote {\bibinfo
  {title} {{Revised constraints and Belle II sensitivity for visible and
  invisible axion-like particles}},}\ }\href {\doibase 10.1007/JHEP12(2017)094}
  {\bibfield  {journal} {\bibinfo  {journal} {JHEP}\ }\textbf {\bibinfo
  {volume} {12}},\ \bibinfo {pages} {094} (\bibinfo {year} {2017})},\ \bibinfo
  {note} {[Erratum: JHEP 03, 190 (2021)]},\ \Eprint
  {http://arxiv.org/abs/1709.00009} {arXiv:1709.00009 [hep-ph]} \BibitemShut
  {NoStop}%
\bibitem [{\citenamefont {D\"obrich}\ \emph {et~al.}(2019)\citenamefont
  {D\"obrich}, \citenamefont {Jaeckel},\ and\ \citenamefont
  {Spadaro}}]{Dobrich:2019dxc}%
  \BibitemOpen
  \bibfield  {author} {\bibinfo {author} {\bibfnamefont {Babette}\ \bibnamefont
  {D\"obrich}}, \bibinfo {author} {\bibfnamefont {Joerg}\ \bibnamefont
  {Jaeckel}}, \ and\ \bibinfo {author} {\bibfnamefont {Tommaso}\ \bibnamefont
  {Spadaro}},\ }\bibfield  {title} {\enquote {\bibinfo {title} {{Light in the
  beam dump - ALP production from decay photons in proton beam-dumps}},}\
  }\href {\doibase 10.1007/JHEP05(2019)213} {\bibfield  {journal} {\bibinfo
  {journal} {JHEP}\ }\textbf {\bibinfo {volume} {05}},\ \bibinfo {pages} {213}
  (\bibinfo {year} {2019})},\ \bibinfo {note} {[Erratum: JHEP 10, 046
  (2020)]},\ \Eprint {http://arxiv.org/abs/1904.02091} {arXiv:1904.02091
  [hep-ph]} \BibitemShut {NoStop}%
\bibitem [{\citenamefont {Harland-Lang}\ \emph {et~al.}(2019)\citenamefont
  {Harland-Lang}, \citenamefont {Jaeckel},\ and\ \citenamefont
  {Spannowsky}}]{Harland-Lang:2019zur}%
  \BibitemOpen
  \bibfield  {author} {\bibinfo {author} {\bibfnamefont {Lucian}\ \bibnamefont
  {Harland-Lang}}, \bibinfo {author} {\bibfnamefont {Joerg}\ \bibnamefont
  {Jaeckel}}, \ and\ \bibinfo {author} {\bibfnamefont {Michael}\ \bibnamefont
  {Spannowsky}},\ }\bibfield  {title} {\enquote {\bibinfo {title} {{A fresh
  look at ALP searches in fixed target experiments}},}\ }\href {\doibase
  10.1016/j.physletb.2019.04.045} {\bibfield  {journal} {\bibinfo  {journal}
  {Phys. Lett. B}\ }\textbf {\bibinfo {volume} {793}},\ \bibinfo {pages}
  {281--289} (\bibinfo {year} {2019})},\ \Eprint
  {http://arxiv.org/abs/1902.04878} {arXiv:1902.04878 [hep-ph]} \BibitemShut
  {NoStop}%
\bibitem [{\citenamefont {Dent}\ \emph {et~al.}(2020)\citenamefont {Dent},
  \citenamefont {Dutta}, \citenamefont {Kim}, \citenamefont {Liao},
  \citenamefont {Mahapatra}, \citenamefont {Sinha},\ and\ \citenamefont
  {Thompson}}]{Dent:2019ueq}%
  \BibitemOpen
  \bibfield  {author} {\bibinfo {author} {\bibfnamefont {James~B.}\
  \bibnamefont {Dent}}, \bibinfo {author} {\bibfnamefont {Bhaskar}\
  \bibnamefont {Dutta}}, \bibinfo {author} {\bibfnamefont {Doojin}\
  \bibnamefont {Kim}}, \bibinfo {author} {\bibfnamefont {Shu}\ \bibnamefont
  {Liao}}, \bibinfo {author} {\bibfnamefont {Rupak}\ \bibnamefont {Mahapatra}},
  \bibinfo {author} {\bibfnamefont {Kuver}\ \bibnamefont {Sinha}}, \ and\
  \bibinfo {author} {\bibfnamefont {Adrian}\ \bibnamefont {Thompson}},\
  }\bibfield  {title} {\enquote {\bibinfo {title} {{New Directions for Axion
  Searches via Scattering at Reactor Neutrino Experiments}},}\ }\href {\doibase
  10.1103/PhysRevLett.124.211804} {\bibfield  {journal} {\bibinfo  {journal}
  {Phys. Rev. Lett.}\ }\textbf {\bibinfo {volume} {124}},\ \bibinfo {pages}
  {211804} (\bibinfo {year} {2020})},\ \Eprint
  {http://arxiv.org/abs/1912.05733} {arXiv:1912.05733 [hep-ph]} \BibitemShut
  {NoStop}%
\bibitem [{\citenamefont {Brdar}\ \emph {et~al.}(2021)\citenamefont {Brdar},
  \citenamefont {Dutta}, \citenamefont {Jang}, \citenamefont {Kim},
  \citenamefont {Shoemaker}, \citenamefont {Tabrizi}, \citenamefont
  {Thompson},\ and\ \citenamefont {Yu}}]{Brdar:2020dpr}%
  \BibitemOpen
  \bibfield  {author} {\bibinfo {author} {\bibfnamefont {Vedran}\ \bibnamefont
  {Brdar}}, \bibinfo {author} {\bibfnamefont {Bhaskar}\ \bibnamefont {Dutta}},
  \bibinfo {author} {\bibfnamefont {Wooyoung}\ \bibnamefont {Jang}}, \bibinfo
  {author} {\bibfnamefont {Doojin}\ \bibnamefont {Kim}}, \bibinfo {author}
  {\bibfnamefont {Ian~M.}\ \bibnamefont {Shoemaker}}, \bibinfo {author}
  {\bibfnamefont {Zahra}\ \bibnamefont {Tabrizi}}, \bibinfo {author}
  {\bibfnamefont {Adrian}\ \bibnamefont {Thompson}}, \ and\ \bibinfo {author}
  {\bibfnamefont {Jaehoon}\ \bibnamefont {Yu}},\ }\bibfield  {title} {\enquote
  {\bibinfo {title} {{Axionlike Particles at Future Neutrino Experiments:
  Closing the Cosmological Triangle}},}\ }\href {\doibase
  10.1103/PhysRevLett.126.201801} {\bibfield  {journal} {\bibinfo  {journal}
  {Phys. Rev. Lett.}\ }\textbf {\bibinfo {volume} {126}},\ \bibinfo {pages}
  {201801} (\bibinfo {year} {2021})},\ \Eprint
  {http://arxiv.org/abs/2011.07054} {arXiv:2011.07054 [hep-ph]} \BibitemShut
  {NoStop}%
\bibitem [{\citenamefont {Jaeckel}\ \emph {et~al.}(2013)\citenamefont
  {Jaeckel}, \citenamefont {Jankowiak},\ and\ \citenamefont
  {Spannowsky}}]{Jaeckel:2012yz}%
  \BibitemOpen
  \bibfield  {author} {\bibinfo {author} {\bibfnamefont {Joerg}\ \bibnamefont
  {Jaeckel}}, \bibinfo {author} {\bibfnamefont {Martin}\ \bibnamefont
  {Jankowiak}}, \ and\ \bibinfo {author} {\bibfnamefont {Michael}\ \bibnamefont
  {Spannowsky}},\ }\bibfield  {title} {\enquote {\bibinfo {title} {{LHC probes
  the hidden sector}},}\ }\href {\doibase 10.1016/j.dark.2013.06.001}
  {\bibfield  {journal} {\bibinfo  {journal} {Phys. Dark Univ.}\ }\textbf
  {\bibinfo {volume} {2}},\ \bibinfo {pages} {111--117} (\bibinfo {year}
  {2013})},\ \Eprint {http://arxiv.org/abs/1212.3620} {arXiv:1212.3620
  [hep-ph]} \BibitemShut {NoStop}%
\bibitem [{\citenamefont {Mimasu}\ and\ \citenamefont
  {Sanz}(2015)}]{Mimasu:2014nea}%
  \BibitemOpen
  \bibfield  {author} {\bibinfo {author} {\bibfnamefont {Ken}\ \bibnamefont
  {Mimasu}}\ and\ \bibinfo {author} {\bibfnamefont {Ver\'onica}\ \bibnamefont
  {Sanz}},\ }\bibfield  {title} {\enquote {\bibinfo {title} {{ALPs at
  Colliders}},}\ }\href {\doibase 10.1007/JHEP06(2015)173} {\bibfield
  {journal} {\bibinfo  {journal} {JHEP}\ }\textbf {\bibinfo {volume} {06}},\
  \bibinfo {pages} {173} (\bibinfo {year} {2015})},\ \Eprint
  {http://arxiv.org/abs/1409.4792} {arXiv:1409.4792 [hep-ph]} \BibitemShut
  {NoStop}%
\bibitem [{\citenamefont {Jaeckel}\ and\ \citenamefont
  {Spannowsky}(2016)}]{Jaeckel:2015jla}%
  \BibitemOpen
  \bibfield  {author} {\bibinfo {author} {\bibfnamefont {Joerg}\ \bibnamefont
  {Jaeckel}}\ and\ \bibinfo {author} {\bibfnamefont {Michael}\ \bibnamefont
  {Spannowsky}},\ }\bibfield  {title} {\enquote {\bibinfo {title} {{Probing MeV
  to 90 GeV axion-like particles with LEP and LHC}},}\ }\href {\doibase
  10.1016/j.physletb.2015.12.037} {\bibfield  {journal} {\bibinfo  {journal}
  {Phys. Lett. B}\ }\textbf {\bibinfo {volume} {753}},\ \bibinfo {pages}
  {482--487} (\bibinfo {year} {2016})},\ \Eprint
  {http://arxiv.org/abs/1509.00476} {arXiv:1509.00476 [hep-ph]} \BibitemShut
  {NoStop}%
\bibitem [{\citenamefont {Izaguirre}\ \emph {et~al.}(2017)\citenamefont
  {Izaguirre}, \citenamefont {Lin},\ and\ \citenamefont
  {Shuve}}]{Izaguirre:2016dfi}%
  \BibitemOpen
  \bibfield  {author} {\bibinfo {author} {\bibfnamefont {Eder}\ \bibnamefont
  {Izaguirre}}, \bibinfo {author} {\bibfnamefont {Tongyan}\ \bibnamefont
  {Lin}}, \ and\ \bibinfo {author} {\bibfnamefont {Brian}\ \bibnamefont
  {Shuve}},\ }\bibfield  {title} {\enquote {\bibinfo {title} {{Searching for
  Axionlike Particles in Flavor-Changing Neutral Current Processes}},}\ }\href
  {\doibase 10.1103/PhysRevLett.118.111802} {\bibfield  {journal} {\bibinfo
  {journal} {Phys. Rev. Lett.}\ }\textbf {\bibinfo {volume} {118}},\ \bibinfo
  {pages} {111802} (\bibinfo {year} {2017})},\ \Eprint
  {http://arxiv.org/abs/1611.09355} {arXiv:1611.09355 [hep-ph]} \BibitemShut
  {NoStop}%
\bibitem [{\citenamefont {Knapen}\ \emph {et~al.}(2017)\citenamefont {Knapen},
  \citenamefont {Lin}, \citenamefont {Lou},\ and\ \citenamefont
  {Melia}}]{Knapen:2016moh}%
  \BibitemOpen
  \bibfield  {author} {\bibinfo {author} {\bibfnamefont {Simon}\ \bibnamefont
  {Knapen}}, \bibinfo {author} {\bibfnamefont {Tongyan}\ \bibnamefont {Lin}},
  \bibinfo {author} {\bibfnamefont {Hou~Keong}\ \bibnamefont {Lou}}, \ and\
  \bibinfo {author} {\bibfnamefont {Tom}\ \bibnamefont {Melia}},\ }\bibfield
  {title} {\enquote {\bibinfo {title} {{Searching for Axionlike Particles with
  Ultraperipheral Heavy-Ion Collisions}},}\ }\href {\doibase
  10.1103/PhysRevLett.118.171801} {\bibfield  {journal} {\bibinfo  {journal}
  {Phys. Rev. Lett.}\ }\textbf {\bibinfo {volume} {118}},\ \bibinfo {pages}
  {171801} (\bibinfo {year} {2017})},\ \Eprint
  {http://arxiv.org/abs/1607.06083} {arXiv:1607.06083 [hep-ph]} \BibitemShut
  {NoStop}%
\bibitem [{\citenamefont {Brivio}\ \emph {et~al.}(2017)\citenamefont {Brivio},
  \citenamefont {Gavela}, \citenamefont {Merlo}, \citenamefont {Mimasu},
  \citenamefont {No}, \citenamefont {del Rey},\ and\ \citenamefont
  {Sanz}}]{Brivio:2017ije}%
  \BibitemOpen
  \bibfield  {author} {\bibinfo {author} {\bibfnamefont {I.}~\bibnamefont
  {Brivio}}, \bibinfo {author} {\bibfnamefont {M.~B.}\ \bibnamefont {Gavela}},
  \bibinfo {author} {\bibfnamefont {L.}~\bibnamefont {Merlo}}, \bibinfo
  {author} {\bibfnamefont {K.}~\bibnamefont {Mimasu}}, \bibinfo {author}
  {\bibfnamefont {J.~M.}\ \bibnamefont {No}}, \bibinfo {author} {\bibfnamefont
  {R.}~\bibnamefont {del Rey}}, \ and\ \bibinfo {author} {\bibfnamefont
  {V.}~\bibnamefont {Sanz}},\ }\bibfield  {title} {\enquote {\bibinfo {title}
  {{ALPs Effective Field Theory and Collider Signatures}},}\ }\href {\doibase
  10.1140/epjc/s10052-017-5111-3} {\bibfield  {journal} {\bibinfo  {journal}
  {Eur. Phys. J. C}\ }\textbf {\bibinfo {volume} {77}},\ \bibinfo {pages} {572}
  (\bibinfo {year} {2017})},\ \Eprint {http://arxiv.org/abs/1701.05379}
  {arXiv:1701.05379 [hep-ph]} \BibitemShut {NoStop}%
\bibitem [{\citenamefont {Mariotti}\ \emph {et~al.}(2018)\citenamefont
  {Mariotti}, \citenamefont {Redigolo}, \citenamefont {Sala},\ and\
  \citenamefont {Tobioka}}]{Mariotti:2017vtv}%
  \BibitemOpen
  \bibfield  {author} {\bibinfo {author} {\bibfnamefont {Alberto}\ \bibnamefont
  {Mariotti}}, \bibinfo {author} {\bibfnamefont {Diego}\ \bibnamefont
  {Redigolo}}, \bibinfo {author} {\bibfnamefont {Filippo}\ \bibnamefont
  {Sala}}, \ and\ \bibinfo {author} {\bibfnamefont {Kohsaku}\ \bibnamefont
  {Tobioka}},\ }\bibfield  {title} {\enquote {\bibinfo {title} {{New LHC bound
  on low-mass diphoton resonances}},}\ }\href {\doibase
  10.1016/j.physletb.2018.06.039} {\bibfield  {journal} {\bibinfo  {journal}
  {Phys. Lett. B}\ }\textbf {\bibinfo {volume} {783}},\ \bibinfo {pages}
  {13--18} (\bibinfo {year} {2018})},\ \Eprint
  {http://arxiv.org/abs/1710.01743} {arXiv:1710.01743 [hep-ph]} \BibitemShut
  {NoStop}%
\bibitem [{\citenamefont {Cid~Vidal}\ \emph {et~al.}(2019)\citenamefont
  {Cid~Vidal}, \citenamefont {Mariotti}, \citenamefont {Redigolo},
  \citenamefont {Sala},\ and\ \citenamefont {Tobioka}}]{CidVidal:2018blh}%
  \BibitemOpen
  \bibfield  {author} {\bibinfo {author} {\bibfnamefont {Xabier}\ \bibnamefont
  {Cid~Vidal}}, \bibinfo {author} {\bibfnamefont {Alberto}\ \bibnamefont
  {Mariotti}}, \bibinfo {author} {\bibfnamefont {Diego}\ \bibnamefont
  {Redigolo}}, \bibinfo {author} {\bibfnamefont {Filippo}\ \bibnamefont
  {Sala}}, \ and\ \bibinfo {author} {\bibfnamefont {Kohsaku}\ \bibnamefont
  {Tobioka}},\ }\bibfield  {title} {\enquote {\bibinfo {title} {{New Axion
  Searches at Flavor Factories}},}\ }\href {\doibase 10.1007/JHEP01(2019)113}
  {\bibfield  {journal} {\bibinfo  {journal} {JHEP}\ }\textbf {\bibinfo
  {volume} {01}},\ \bibinfo {pages} {113} (\bibinfo {year} {2019})},\ \bibinfo
  {note} {[Erratum: JHEP 06, 141 (2020)]},\ \Eprint
  {http://arxiv.org/abs/1810.09452} {arXiv:1810.09452 [hep-ph]} \BibitemShut
  {NoStop}%
\bibitem [{\citenamefont {Beacham}\ \emph {et~al.}(2020)\citenamefont {Beacham}
  \emph {et~al.}}]{Beacham:2019nyx}%
  \BibitemOpen
  \bibfield  {author} {\bibinfo {author} {\bibfnamefont {J.}~\bibnamefont
  {Beacham}} \emph {et~al.},\ }\bibfield  {title} {\enquote {\bibinfo {title}
  {{Physics Beyond Colliders at CERN: Beyond the Standard Model Working Group
  Report}},}\ }\href {\doibase 10.1088/1361-6471/ab4cd2} {\bibfield  {journal}
  {\bibinfo  {journal} {J. Phys. G}\ }\textbf {\bibinfo {volume} {47}},\
  \bibinfo {pages} {010501} (\bibinfo {year} {2020})},\ \Eprint
  {http://arxiv.org/abs/1901.09966} {arXiv:1901.09966 [hep-ex]} \BibitemShut
  {NoStop}%
\bibitem [{\citenamefont {Alonso-\'Alvarez}\ \emph {et~al.}(2019)\citenamefont
  {Alonso-\'Alvarez}, \citenamefont {Gavela},\ and\ \citenamefont
  {Quilez}}]{Alonso-Alvarez:2018irt}%
  \BibitemOpen
  \bibfield  {author} {\bibinfo {author} {\bibfnamefont {G.}~\bibnamefont
  {Alonso-\'Alvarez}}, \bibinfo {author} {\bibfnamefont {M.~B.}\ \bibnamefont
  {Gavela}}, \ and\ \bibinfo {author} {\bibfnamefont {P.}~\bibnamefont
  {Quilez}},\ }\bibfield  {title} {\enquote {\bibinfo {title} {{Axion couplings
  to electroweak gauge bosons}},}\ }\href {\doibase
  10.1140/epjc/s10052-019-6732-5} {\bibfield  {journal} {\bibinfo  {journal}
  {Eur. Phys. J. C}\ }\textbf {\bibinfo {volume} {79}},\ \bibinfo {pages} {223}
  (\bibinfo {year} {2019})},\ \Eprint {http://arxiv.org/abs/1811.05466}
  {arXiv:1811.05466 [hep-ph]} \BibitemShut {NoStop}%
\bibitem [{\citenamefont {Ebadi}\ \emph {et~al.}(2019)\citenamefont {Ebadi},
  \citenamefont {Khatibi},\ and\ \citenamefont
  {Mohammadi~Najafabadi}}]{Ebadi:2019gij}%
  \BibitemOpen
  \bibfield  {author} {\bibinfo {author} {\bibfnamefont {Javad}\ \bibnamefont
  {Ebadi}}, \bibinfo {author} {\bibfnamefont {Sara}\ \bibnamefont {Khatibi}}, \
  and\ \bibinfo {author} {\bibfnamefont {Mojtaba}\ \bibnamefont
  {Mohammadi~Najafabadi}},\ }\bibfield  {title} {\enquote {\bibinfo {title}
  {{New probes for axionlike particles at hadron colliders}},}\ }\href
  {\doibase 10.1103/PhysRevD.100.015016} {\bibfield  {journal} {\bibinfo
  {journal} {Phys. Rev. D}\ }\textbf {\bibinfo {volume} {100}},\ \bibinfo
  {pages} {015016} (\bibinfo {year} {2019})},\ \Eprint
  {http://arxiv.org/abs/1901.03061} {arXiv:1901.03061 [hep-ph]} \BibitemShut
  {NoStop}%
\bibitem [{\citenamefont {Gavela}\ \emph {et~al.}(2020)\citenamefont {Gavela},
  \citenamefont {No}, \citenamefont {Sanz},\ and\ \citenamefont
  {de~Troc\'oniz}}]{Gavela:2019cmq}%
  \BibitemOpen
  \bibfield  {author} {\bibinfo {author} {\bibfnamefont {M.~B.}\ \bibnamefont
  {Gavela}}, \bibinfo {author} {\bibfnamefont {J.~M.}\ \bibnamefont {No}},
  \bibinfo {author} {\bibfnamefont {V.}~\bibnamefont {Sanz}}, \ and\ \bibinfo
  {author} {\bibfnamefont {J.~F.}\ \bibnamefont {de~Troc\'oniz}},\ }\bibfield
  {title} {\enquote {\bibinfo {title} {{Nonresonant Searches for Axionlike
  Particles at the LHC}},}\ }\href {\doibase 10.1103/PhysRevLett.124.051802}
  {\bibfield  {journal} {\bibinfo  {journal} {Phys. Rev. Lett.}\ }\textbf
  {\bibinfo {volume} {124}},\ \bibinfo {pages} {051802} (\bibinfo {year}
  {2020})},\ \Eprint {http://arxiv.org/abs/1905.12953} {arXiv:1905.12953
  [hep-ph]} \BibitemShut {NoStop}%
\bibitem [{\citenamefont {Altmannshofer}\ \emph {et~al.}(2020)\citenamefont
  {Altmannshofer}, \citenamefont {Gori},\ and\ \citenamefont
  {Robinson}}]{Altmannshofer:2019yji}%
  \BibitemOpen
  \bibfield  {author} {\bibinfo {author} {\bibfnamefont {Wolfgang}\
  \bibnamefont {Altmannshofer}}, \bibinfo {author} {\bibfnamefont {Stefania}\
  \bibnamefont {Gori}}, \ and\ \bibinfo {author} {\bibfnamefont {Dean~J.}\
  \bibnamefont {Robinson}},\ }\bibfield  {title} {\enquote {\bibinfo {title}
  {{Constraining axionlike particles from rare pion decays}},}\ }\href
  {\doibase 10.1103/PhysRevD.101.075002} {\bibfield  {journal} {\bibinfo
  {journal} {Phys. Rev. D}\ }\textbf {\bibinfo {volume} {101}},\ \bibinfo
  {pages} {075002} (\bibinfo {year} {2020})},\ \Eprint
  {http://arxiv.org/abs/1909.00005} {arXiv:1909.00005 [hep-ph]} \BibitemShut
  {NoStop}%
\bibitem [{\citenamefont {Gershtein}\ \emph {et~al.}(2021)\citenamefont
  {Gershtein}, \citenamefont {Knapen},\ and\ \citenamefont
  {Redigolo}}]{Gershtein:2020mwi}%
  \BibitemOpen
  \bibfield  {author} {\bibinfo {author} {\bibfnamefont {Yuri}\ \bibnamefont
  {Gershtein}}, \bibinfo {author} {\bibfnamefont {Simon}\ \bibnamefont
  {Knapen}}, \ and\ \bibinfo {author} {\bibfnamefont {Diego}\ \bibnamefont
  {Redigolo}},\ }\bibfield  {title} {\enquote {\bibinfo {title} {{Probing
  naturally light singlets with a displaced vertex trigger}},}\ }\href
  {\doibase 10.1016/j.physletb.2021.136758} {\bibfield  {journal} {\bibinfo
  {journal} {Phys. Lett. B}\ }\textbf {\bibinfo {volume} {823}},\ \bibinfo
  {pages} {136758} (\bibinfo {year} {2021})},\ \Eprint
  {http://arxiv.org/abs/2012.07864} {arXiv:2012.07864 [hep-ph]} \BibitemShut
  {NoStop}%
\bibitem [{\citenamefont {Di~Luzio}\ \emph {et~al.}(2020)\citenamefont
  {Di~Luzio}, \citenamefont {Giannotti}, \citenamefont {Nardi},\ and\
  \citenamefont {Visinelli}}]{DiLuzio:2020wdo}%
  \BibitemOpen
  \bibfield  {author} {\bibinfo {author} {\bibfnamefont {Luca}\ \bibnamefont
  {Di~Luzio}}, \bibinfo {author} {\bibfnamefont {Maurizio}\ \bibnamefont
  {Giannotti}}, \bibinfo {author} {\bibfnamefont {Enrico}\ \bibnamefont
  {Nardi}}, \ and\ \bibinfo {author} {\bibfnamefont {Luca}\ \bibnamefont
  {Visinelli}},\ }\bibfield  {title} {\enquote {\bibinfo {title} {{The
  landscape of QCD axion models}},}\ }\href {\doibase
  10.1016/j.physrep.2020.06.002} {\bibfield  {journal} {\bibinfo  {journal}
  {Phys. Rept.}\ }\textbf {\bibinfo {volume} {870}},\ \bibinfo {pages} {1--117}
  (\bibinfo {year} {2020})},\ \Eprint {http://arxiv.org/abs/2003.01100}
  {arXiv:2003.01100 [hep-ph]} \BibitemShut {NoStop}%
\bibitem [{\citenamefont {Knapen}\ \emph {et~al.}(2022)\citenamefont {Knapen},
  \citenamefont {Kumar},\ and\ \citenamefont {Redigolo}}]{Knapen:2021elo}%
  \BibitemOpen
  \bibfield  {author} {\bibinfo {author} {\bibfnamefont {Simon}\ \bibnamefont
  {Knapen}}, \bibinfo {author} {\bibfnamefont {Soubhik}\ \bibnamefont {Kumar}},
  \ and\ \bibinfo {author} {\bibfnamefont {Diego}\ \bibnamefont {Redigolo}},\
  }\bibfield  {title} {\enquote {\bibinfo {title} {{Searching for axionlike
  particles with data scouting at ATLAS and CMS}},}\ }\href {\doibase
  10.1103/PhysRevD.105.115012} {\bibfield  {journal} {\bibinfo  {journal}
  {Phys. Rev. D}\ }\textbf {\bibinfo {volume} {105}},\ \bibinfo {pages}
  {115012} (\bibinfo {year} {2022})},\ \Eprint
  {http://arxiv.org/abs/2112.07720} {arXiv:2112.07720 [hep-ph]} \BibitemShut
  {NoStop}%
\bibitem [{\citenamefont {Co}\ \emph {et~al.}(2023)\citenamefont {Co},
  \citenamefont {Kumar},\ and\ \citenamefont {Liu}}]{Co:2022bqq}%
  \BibitemOpen
  \bibfield  {author} {\bibinfo {author} {\bibfnamefont {Raymond~T.}\
  \bibnamefont {Co}}, \bibinfo {author} {\bibfnamefont {Soubhik}\ \bibnamefont
  {Kumar}}, \ and\ \bibinfo {author} {\bibfnamefont {Zhen}\ \bibnamefont
  {Liu}},\ }\bibfield  {title} {\enquote {\bibinfo {title} {{Searches for heavy
  QCD axions via dimuon final states}},}\ }\href {\doibase
  10.1007/JHEP02(2023)111} {\bibfield  {journal} {\bibinfo  {journal} {JHEP}\
  }\textbf {\bibinfo {volume} {02}},\ \bibinfo {pages} {111} (\bibinfo {year}
  {2023})},\ \Eprint {http://arxiv.org/abs/2210.02462} {arXiv:2210.02462
  [hep-ph]} \BibitemShut {NoStop}%
\bibitem [{\citenamefont {Acciarri}\ \emph {et~al.}(2023)\citenamefont
  {Acciarri} \emph {et~al.}}]{ArgoNeuT:2022mrm}%
  \BibitemOpen
  \bibfield  {author} {\bibinfo {author} {\bibfnamefont {R.}~\bibnamefont
  {Acciarri}} \emph {et~al.} (\bibinfo {collaboration} {ArgoNeuT}),\ }\bibfield
   {title} {\enquote {\bibinfo {title} {{First Constraints on Heavy QCD Axions
  with a Liquid Argon Time Projection Chamber Using the ArgoNeuT
  Experiment}},}\ }\href {\doibase 10.1103/PhysRevLett.130.221802} {\bibfield
  {journal} {\bibinfo  {journal} {Phys. Rev. Lett.}\ }\textbf {\bibinfo
  {volume} {130}},\ \bibinfo {pages} {221802} (\bibinfo {year} {2023})},\
  \Eprint {http://arxiv.org/abs/2207.08448} {arXiv:2207.08448 [hep-ex]}
  \BibitemShut {NoStop}%
\bibitem [{\citenamefont {Coloma}\ \emph {et~al.}(2022)\citenamefont {Coloma},
  \citenamefont {Hern\'andez},\ and\ \citenamefont {Urrea}}]{Coloma:2022hlv}%
  \BibitemOpen
  \bibfield  {author} {\bibinfo {author} {\bibfnamefont {Pilar}\ \bibnamefont
  {Coloma}}, \bibinfo {author} {\bibfnamefont {Pilar}\ \bibnamefont
  {Hern\'andez}}, \ and\ \bibinfo {author} {\bibfnamefont {Salvador}\
  \bibnamefont {Urrea}},\ }\bibfield  {title} {\enquote {\bibinfo {title} {{New
  bounds on axion-like particles from MicroBooNE}},}\ }\href {\doibase
  10.1007/JHEP08(2022)025} {\bibfield  {journal} {\bibinfo  {journal} {JHEP}\
  }\textbf {\bibinfo {volume} {08}},\ \bibinfo {pages} {025} (\bibinfo {year}
  {2022})},\ \Eprint {http://arxiv.org/abs/2202.03447} {arXiv:2202.03447
  [hep-ph]} \BibitemShut {NoStop}%
\bibitem [{\citenamefont {Batell}\ \emph {et~al.}(2019)\citenamefont {Batell},
  \citenamefont {Berger},\ and\ \citenamefont {Ismail}}]{Batell:2019nwo}%
  \BibitemOpen
  \bibfield  {author} {\bibinfo {author} {\bibfnamefont {Brian}\ \bibnamefont
  {Batell}}, \bibinfo {author} {\bibfnamefont {Joshua}\ \bibnamefont {Berger}},
  \ and\ \bibinfo {author} {\bibfnamefont {Ahmed}\ \bibnamefont {Ismail}},\
  }\bibfield  {title} {\enquote {\bibinfo {title} {{Probing the Higgs Portal at
  the Fermilab Short-Baseline Neutrino Experiments}},}\ }\href {\doibase
  10.1103/PhysRevD.100.115039} {\bibfield  {journal} {\bibinfo  {journal}
  {Phys. Rev. D}\ }\textbf {\bibinfo {volume} {100}},\ \bibinfo {pages}
  {115039} (\bibinfo {year} {2019})},\ \Eprint
  {http://arxiv.org/abs/1909.11670} {arXiv:1909.11670 [hep-ph]} \BibitemShut
  {NoStop}%
\bibitem [{\citenamefont {Spitz}(2014)}]{Spitz:2014hwa}%
  \BibitemOpen
  \bibfield  {author} {\bibinfo {author} {\bibfnamefont {J.}~\bibnamefont
  {Spitz}},\ }\bibfield  {title} {\enquote {\bibinfo {title} {{Cross Section
  Measurements with Monoenergetic Muon Neutrinos}},}\ }\href {\doibase
  10.1103/PhysRevD.89.073007} {\bibfield  {journal} {\bibinfo  {journal} {Phys.
  Rev. D}\ }\textbf {\bibinfo {volume} {89}},\ \bibinfo {pages} {073007}
  (\bibinfo {year} {2014})},\ \Eprint {http://arxiv.org/abs/1402.2284}
  {arXiv:1402.2284 [physics.ins-det]} \BibitemShut {NoStop}%
\bibitem [{\citenamefont {Nikolakopoulos}\ \emph {et~al.}(2021)\citenamefont
  {Nikolakopoulos}, \citenamefont {Pandey}, \citenamefont {Spitz},\ and\
  \citenamefont {Jachowicz}}]{Nikolakopoulos:2020alk}%
  \BibitemOpen
  \bibfield  {author} {\bibinfo {author} {\bibfnamefont {A.}~\bibnamefont
  {Nikolakopoulos}}, \bibinfo {author} {\bibfnamefont {V.}~\bibnamefont
  {Pandey}}, \bibinfo {author} {\bibfnamefont {J.}~\bibnamefont {Spitz}}, \
  and\ \bibinfo {author} {\bibfnamefont {Natalie}\ \bibnamefont {Jachowicz}},\
  }\bibfield  {title} {\enquote {\bibinfo {title} {{Modeling quasielastic
  interactions of monoenergetic kaon decay-at-rest neutrinos}},}\ }\href
  {\doibase 10.1103/PhysRevC.103.064603} {\bibfield  {journal} {\bibinfo
  {journal} {Phys. Rev. C}\ }\textbf {\bibinfo {volume} {103}},\ \bibinfo
  {pages} {064603} (\bibinfo {year} {2021})},\ \Eprint
  {http://arxiv.org/abs/2010.05794} {arXiv:2010.05794 [nucl-th]} \BibitemShut
  {NoStop}%
\bibitem [{\citenamefont {Kelly}\ \emph {et~al.}(2021)\citenamefont {Kelly},
  \citenamefont {Kumar},\ and\ \citenamefont {Liu}}]{Kelly:2020dda}%
  \BibitemOpen
  \bibfield  {author} {\bibinfo {author} {\bibfnamefont {Kevin~J.}\
  \bibnamefont {Kelly}}, \bibinfo {author} {\bibfnamefont {Soubhik}\
  \bibnamefont {Kumar}}, \ and\ \bibinfo {author} {\bibfnamefont {Zhen}\
  \bibnamefont {Liu}},\ }\bibfield  {title} {\enquote {\bibinfo {title} {{Heavy
  axion opportunities at the DUNE near detector}},}\ }\href {\doibase
  10.1103/PhysRevD.103.095002} {\bibfield  {journal} {\bibinfo  {journal}
  {Phys. Rev. D}\ }\textbf {\bibinfo {volume} {103}},\ \bibinfo {pages}
  {095002} (\bibinfo {year} {2021})},\ \Eprint
  {http://arxiv.org/abs/2011.05995} {arXiv:2011.05995 [hep-ph]} \BibitemShut
  {NoStop}%
\bibitem [{\citenamefont {Jerhot}\ \emph {et~al.}(2022)\citenamefont {Jerhot},
  \citenamefont {D\"obrich}, \citenamefont {Ertas}, \citenamefont
  {Kahlhoefer},\ and\ \citenamefont {Spadaro}}]{Jerhot:2022chi}%
  \BibitemOpen
  \bibfield  {author} {\bibinfo {author} {\bibfnamefont {Jan}\ \bibnamefont
  {Jerhot}}, \bibinfo {author} {\bibfnamefont {Babette}\ \bibnamefont
  {D\"obrich}}, \bibinfo {author} {\bibfnamefont {Fatih}\ \bibnamefont
  {Ertas}}, \bibinfo {author} {\bibfnamefont {Felix}\ \bibnamefont
  {Kahlhoefer}}, \ and\ \bibinfo {author} {\bibfnamefont {Tommaso}\
  \bibnamefont {Spadaro}},\ }\bibfield  {title} {\enquote {\bibinfo {title}
  {{ALPINIST: Axion-Like Particles In Numerous Interactions Simulated and
  Tabulated}},}\ }\href {\doibase 10.1007/JHEP07(2022)094} {\bibfield
  {journal} {\bibinfo  {journal} {JHEP}\ }\textbf {\bibinfo {volume} {07}},\
  \bibinfo {pages} {094} (\bibinfo {year} {2022})},\ \Eprint
  {http://arxiv.org/abs/2201.05170} {arXiv:2201.05170 [hep-ph]} \BibitemShut
  {NoStop}%
\bibitem [{\citenamefont {Afik}\ \emph {et~al.}(2023)\citenamefont {Afik},
  \citenamefont {D\"obrich}, \citenamefont {Jerhot}, \citenamefont {Soreq},\
  and\ \citenamefont {Tobioka}}]{Afik:2023mhj}%
  \BibitemOpen
  \bibfield  {author} {\bibinfo {author} {\bibfnamefont {Yoav}\ \bibnamefont
  {Afik}}, \bibinfo {author} {\bibfnamefont {Babette}\ \bibnamefont
  {D\"obrich}}, \bibinfo {author} {\bibfnamefont {Jan}\ \bibnamefont {Jerhot}},
  \bibinfo {author} {\bibfnamefont {Yotam}\ \bibnamefont {Soreq}}, \ and\
  \bibinfo {author} {\bibfnamefont {Kohsaku}\ \bibnamefont {Tobioka}},\
  }\bibfield  {title} {\enquote {\bibinfo {title} {{Probing Long-lived Axions
  at the KOTO Experiment}},}\ }\href@noop {} {\  (\bibinfo {year} {2023})},\
  \Eprint {http://arxiv.org/abs/2303.01521} {arXiv:2303.01521 [hep-ph]}
  \BibitemShut {NoStop}%
\bibitem [{\citenamefont {Bauer}\ \emph
  {et~al.}(2021{\natexlab{a}})\citenamefont {Bauer}, \citenamefont {Neubert},
  \citenamefont {Renner}, \citenamefont {Schnubel},\ and\ \citenamefont
  {Thamm}}]{Bauer:2021wjo}%
  \BibitemOpen
  \bibfield  {author} {\bibinfo {author} {\bibfnamefont {Martin}\ \bibnamefont
  {Bauer}}, \bibinfo {author} {\bibfnamefont {Matthias}\ \bibnamefont
  {Neubert}}, \bibinfo {author} {\bibfnamefont {Sophie}\ \bibnamefont
  {Renner}}, \bibinfo {author} {\bibfnamefont {Marvin}\ \bibnamefont
  {Schnubel}}, \ and\ \bibinfo {author} {\bibfnamefont {Andrea}\ \bibnamefont
  {Thamm}},\ }\bibfield  {title} {\enquote {\bibinfo {title} {{Consistent
  Treatment of Axions in the Weak Chiral Lagrangian}},}\ }\href {\doibase
  10.1103/PhysRevLett.127.081803} {\bibfield  {journal} {\bibinfo  {journal}
  {Phys. Rev. Lett.}\ }\textbf {\bibinfo {volume} {127}},\ \bibinfo {pages}
  {081803} (\bibinfo {year} {2021}{\natexlab{a}})},\ \Eprint
  {http://arxiv.org/abs/2102.13112} {arXiv:2102.13112 [hep-ph]} \BibitemShut
  {NoStop}%
\bibitem [{\citenamefont {Abratenko}\ \emph {et~al.}(2021)\citenamefont
  {Abratenko} \emph {et~al.}}]{MicroBooNE:2021usw}%
  \BibitemOpen
  \bibfield  {author} {\bibinfo {author} {\bibfnamefont {P.}~\bibnamefont
  {Abratenko}} \emph {et~al.} (\bibinfo {collaboration} {MicroBooNE}),\
  }\bibfield  {title} {\enquote {\bibinfo {title} {{Search for a Higgs Portal
  Scalar Decaying to Electron-Positron Pairs in the MicroBooNE Detector}},}\
  }\href {\doibase 10.1103/PhysRevLett.127.151803} {\bibfield  {journal}
  {\bibinfo  {journal} {Phys. Rev. Lett.}\ }\textbf {\bibinfo {volume} {127}},\
  \bibinfo {pages} {151803} (\bibinfo {year} {2021})},\ \Eprint
  {http://arxiv.org/abs/2106.00568} {arXiv:2106.00568 [hep-ex]} \BibitemShut
  {NoStop}%
\bibitem [{\citenamefont {Chang}\ \emph {et~al.}(2018)\citenamefont {Chang},
  \citenamefont {Essig},\ and\ \citenamefont {McDermott}}]{Chang:2018rso}%
  \BibitemOpen
  \bibfield  {author} {\bibinfo {author} {\bibfnamefont {Jae~Hyeok}\
  \bibnamefont {Chang}}, \bibinfo {author} {\bibfnamefont {Rouven}\
  \bibnamefont {Essig}}, \ and\ \bibinfo {author} {\bibfnamefont {Samuel~D.}\
  \bibnamefont {McDermott}},\ }\bibfield  {title} {\enquote {\bibinfo {title}
  {{Supernova 1987A Constraints on Sub-GeV Dark Sectors, Millicharged
  Particles, the QCD Axion, and an Axion-like Particle}},}\ }\href {\doibase
  10.1007/JHEP09(2018)051} {\bibfield  {journal} {\bibinfo  {journal} {JHEP}\
  }\textbf {\bibinfo {volume} {09}},\ \bibinfo {pages} {051} (\bibinfo {year}
  {2018})},\ \Eprint {http://arxiv.org/abs/1803.00993} {arXiv:1803.00993
  [hep-ph]} \BibitemShut {NoStop}%
\bibitem [{\citenamefont {Ertas}\ and\ \citenamefont
  {Kahlhoefer}(2020)}]{Ertas:2020xcc}%
  \BibitemOpen
  \bibfield  {author} {\bibinfo {author} {\bibfnamefont {Fatih}\ \bibnamefont
  {Ertas}}\ and\ \bibinfo {author} {\bibfnamefont {Felix}\ \bibnamefont
  {Kahlhoefer}},\ }\bibfield  {title} {\enquote {\bibinfo {title} {{On the
  interplay between astrophysical and laboratory probes of MeV-scale axion-like
  particles}},}\ }\href {\doibase 10.1007/JHEP07(2020)050} {\bibfield
  {journal} {\bibinfo  {journal} {JHEP}\ }\textbf {\bibinfo {volume} {07}},\
  \bibinfo {pages} {050} (\bibinfo {year} {2020})},\ \Eprint
  {http://arxiv.org/abs/2004.01193} {arXiv:2004.01193 [hep-ph]} \BibitemShut
  {NoStop}%
\bibitem [{\citenamefont {Depta}\ \emph {et~al.}(2020)\citenamefont {Depta},
  \citenamefont {Hufnagel},\ and\ \citenamefont
  {Schmidt-Hoberg}}]{Depta:2020wmr}%
  \BibitemOpen
  \bibfield  {author} {\bibinfo {author} {\bibfnamefont {Paul~Frederik}\
  \bibnamefont {Depta}}, \bibinfo {author} {\bibfnamefont {Marco}\ \bibnamefont
  {Hufnagel}}, \ and\ \bibinfo {author} {\bibfnamefont {Kai}\ \bibnamefont
  {Schmidt-Hoberg}},\ }\bibfield  {title} {\enquote {\bibinfo {title} {{Robust
  cosmological constraints on axion-like particles}},}\ }\href {\doibase
  10.1088/1475-7516/2020/05/009} {\bibfield  {journal} {\bibinfo  {journal}
  {JCAP}\ }\textbf {\bibinfo {volume} {05}},\ \bibinfo {pages} {009} (\bibinfo
  {year} {2020})},\ \Eprint {http://arxiv.org/abs/2002.08370} {arXiv:2002.08370
  [hep-ph]} \BibitemShut {NoStop}%
\bibitem [{\citenamefont {Artamonov}\ \emph {et~al.}(2009)\citenamefont
  {Artamonov} \emph {et~al.}}]{BNL-E949:2009dza}%
  \BibitemOpen
  \bibfield  {author} {\bibinfo {author} {\bibfnamefont {A.~V.}\ \bibnamefont
  {Artamonov}} \emph {et~al.} (\bibinfo {collaboration} {BNL-E949}),\
  }\bibfield  {title} {\enquote {\bibinfo {title} {{Study of the decay
  $K^+\to\pi^+\nu \bar\nu$ in the momentum region $140 < P_\pi < 199$
  MeV/c}},}\ }\href {\doibase 10.1103/PhysRevD.79.092004} {\bibfield  {journal}
  {\bibinfo  {journal} {Phys. Rev. D}\ }\textbf {\bibinfo {volume} {79}},\
  \bibinfo {pages} {092004} (\bibinfo {year} {2009})},\ \Eprint
  {http://arxiv.org/abs/0903.0030} {arXiv:0903.0030 [hep-ex]} \BibitemShut
  {NoStop}%
\bibitem [{\citenamefont {Cortina~Gil}\ \emph {et~al.}(2021)\citenamefont
  {Cortina~Gil} \emph {et~al.}}]{NA62:2021zjw}%
  \BibitemOpen
  \bibfield  {author} {\bibinfo {author} {\bibfnamefont {Eduardo}\ \bibnamefont
  {Cortina~Gil}} \emph {et~al.} (\bibinfo {collaboration} {NA62}),\ }\bibfield
  {title} {\enquote {\bibinfo {title} {{Measurement of the very rare
  K$^{+}$\textrightarrow{}$ {\pi}^{+}\nu \overline{\nu} $ decay}},}\ }\href
  {\doibase 10.1007/JHEP06(2021)093} {\bibfield  {journal} {\bibinfo  {journal}
  {JHEP}\ }\textbf {\bibinfo {volume} {06}},\ \bibinfo {pages} {093} (\bibinfo
  {year} {2021})},\ \Eprint {http://arxiv.org/abs/2103.15389} {arXiv:2103.15389
  [hep-ex]} \BibitemShut {NoStop}%
\bibitem [{\citenamefont {Bergsma}\ \emph {et~al.}(1985)\citenamefont {Bergsma}
  \emph {et~al.}}]{CHARM:1985anb}%
  \BibitemOpen
  \bibfield  {author} {\bibinfo {author} {\bibfnamefont {F.}~\bibnamefont
  {Bergsma}} \emph {et~al.} (\bibinfo {collaboration} {CHARM}),\ }\bibfield
  {title} {\enquote {\bibinfo {title} {{Search for Axion Like Particle
  Production in 400-{GeV} Proton - Copper Interactions}},}\ }\href {\doibase
  10.1016/0370-2693(85)90400-9} {\bibfield  {journal} {\bibinfo  {journal}
  {Phys. Lett. B}\ }\textbf {\bibinfo {volume} {157}},\ \bibinfo {pages}
  {458--462} (\bibinfo {year} {1985})}\BibitemShut {NoStop}%
\bibitem [{\citenamefont {Blumlein}\ \emph {et~al.}(1991)\citenamefont
  {Blumlein} \emph {et~al.}}]{Blumlein:1990ay}%
  \BibitemOpen
  \bibfield  {author} {\bibinfo {author} {\bibfnamefont {J.}~\bibnamefont
  {Blumlein}} \emph {et~al.},\ }\bibfield  {title} {\enquote {\bibinfo {title}
  {{Limits on neutral light scalar and pseudoscalar particles in a proton beam
  dump experiment}},}\ }\href {\doibase 10.1007/BF01548556} {\bibfield
  {journal} {\bibinfo  {journal} {Z. Phys. C}\ }\textbf {\bibinfo {volume}
  {51}},\ \bibinfo {pages} {341--350} (\bibinfo {year} {1991})}\BibitemShut
  {NoStop}%
\bibitem [{\citenamefont {Aidala}\ \emph {et~al.}(2019)\citenamefont {Aidala}
  \emph {et~al.}}]{SeaQuest:2017kjt}%
  \BibitemOpen
  \bibfield  {author} {\bibinfo {author} {\bibfnamefont {C.~A.}\ \bibnamefont
  {Aidala}} \emph {et~al.} (\bibinfo {collaboration} {SeaQuest}),\ }\bibfield
  {title} {\enquote {\bibinfo {title} {{The SeaQuest Spectrometer at
  Fermilab}},}\ }\href {\doibase 10.1016/j.nima.2019.03.039} {\bibfield
  {journal} {\bibinfo  {journal} {Nucl. Instrum. Meth. A}\ }\textbf {\bibinfo
  {volume} {930}},\ \bibinfo {pages} {49--63} (\bibinfo {year} {2019})},\
  \Eprint {http://arxiv.org/abs/1706.09990} {arXiv:1706.09990
  [physics.ins-det]} \BibitemShut {NoStop}%
\bibitem [{\citenamefont {Berlin}\ \emph {et~al.}(2018)\citenamefont {Berlin},
  \citenamefont {Gori}, \citenamefont {Schuster},\ and\ \citenamefont
  {Toro}}]{Berlin:2018pwi}%
  \BibitemOpen
  \bibfield  {author} {\bibinfo {author} {\bibfnamefont {Asher}\ \bibnamefont
  {Berlin}}, \bibinfo {author} {\bibfnamefont {Stefania}\ \bibnamefont {Gori}},
  \bibinfo {author} {\bibfnamefont {Philip}\ \bibnamefont {Schuster}}, \ and\
  \bibinfo {author} {\bibfnamefont {Natalia}\ \bibnamefont {Toro}},\ }\bibfield
   {title} {\enquote {\bibinfo {title} {{Dark Sectors at the Fermilab SeaQuest
  Experiment}},}\ }\href {\doibase 10.1103/PhysRevD.98.035011} {\bibfield
  {journal} {\bibinfo  {journal} {Phys. Rev. D}\ }\textbf {\bibinfo {volume}
  {98}},\ \bibinfo {pages} {035011} (\bibinfo {year} {2018})},\ \Eprint
  {http://arxiv.org/abs/1804.00661} {arXiv:1804.00661 [hep-ph]} \BibitemShut
  {NoStop}%
\bibitem [{\citenamefont {Blinov}\ \emph {et~al.}(2022)\citenamefont {Blinov},
  \citenamefont {Kowalczyk},\ and\ \citenamefont {Wynne}}]{Blinov:2021say}%
  \BibitemOpen
  \bibfield  {author} {\bibinfo {author} {\bibfnamefont {Nikita}\ \bibnamefont
  {Blinov}}, \bibinfo {author} {\bibfnamefont {Elizabeth}\ \bibnamefont
  {Kowalczyk}}, \ and\ \bibinfo {author} {\bibfnamefont {Margaret}\
  \bibnamefont {Wynne}},\ }\bibfield  {title} {\enquote {\bibinfo {title}
  {{Axion-like particle searches at DarkQuest}},}\ }\href {\doibase
  10.1007/JHEP02(2022)036} {\bibfield  {journal} {\bibinfo  {journal} {JHEP}\
  }\textbf {\bibinfo {volume} {02}},\ \bibinfo {pages} {036} (\bibinfo {year}
  {2022})},\ \Eprint {http://arxiv.org/abs/2112.09814} {arXiv:2112.09814
  [hep-ph]} \BibitemShut {NoStop}%
\bibitem [{\citenamefont {Ariga}\ \emph {et~al.}(2019)\citenamefont {Ariga}
  \emph {et~al.}}]{FASER:2018eoc}%
  \BibitemOpen
  \bibfield  {author} {\bibinfo {author} {\bibfnamefont {Akitaka}\ \bibnamefont
  {Ariga}} \emph {et~al.} (\bibinfo {collaboration} {FASER}),\ }\bibfield
  {title} {\enquote {\bibinfo {title} {{FASER\textquoteright{}s physics reach
  for long-lived particles}},}\ }\href {\doibase 10.1103/PhysRevD.99.095011}
  {\bibfield  {journal} {\bibinfo  {journal} {Phys. Rev. D}\ }\textbf {\bibinfo
  {volume} {99}},\ \bibinfo {pages} {095011} (\bibinfo {year} {2019})},\
  \Eprint {http://arxiv.org/abs/1811.12522} {arXiv:1811.12522 [hep-ph]}
  \BibitemShut {NoStop}%
\bibitem [{\citenamefont {Yamanaka}(2012)}]{Yamanaka:2012yma}%
  \BibitemOpen
  \bibfield  {author} {\bibinfo {author} {\bibfnamefont {Taku}\ \bibnamefont
  {Yamanaka}} (\bibinfo {collaboration} {KOTO}),\ }\bibfield  {title} {\enquote
  {\bibinfo {title} {{The J-PARC KOTO experiment}},}\ }\href {\doibase
  10.1093/ptep/pts057} {\bibfield  {journal} {\bibinfo  {journal} {PTEP}\
  }\textbf {\bibinfo {volume} {2012}},\ \bibinfo {pages} {02B006} (\bibinfo
  {year} {2012})}\BibitemShut {NoStop}%
\bibitem [{\citenamefont {Anelli}\ \emph {et~al.}(2015)\citenamefont {Anelli}
  \emph {et~al.}}]{SHiP:2015vad}%
  \BibitemOpen
  \bibfield  {author} {\bibinfo {author} {\bibfnamefont {M.}~\bibnamefont
  {Anelli}} \emph {et~al.} (\bibinfo {collaboration} {SHiP}),\ }\bibfield
  {title} {\enquote {\bibinfo {title} {{A facility to Search for Hidden
  Particles (SHiP) at the CERN SPS}},}\ }\href@noop {} {\  (\bibinfo {year}
  {2015})},\ \Eprint {http://arxiv.org/abs/1504.04956} {arXiv:1504.04956
  [physics.ins-det]} \BibitemShut {NoStop}%
\bibitem [{\citenamefont {Bauer}\ \emph
  {et~al.}(2021{\natexlab{b}})\citenamefont {Bauer}, \citenamefont {Neubert},
  \citenamefont {Renner}, \citenamefont {Schnubel},\ and\ \citenamefont
  {Thamm}}]{Bauer:2020jbp}%
  \BibitemOpen
  \bibfield  {author} {\bibinfo {author} {\bibfnamefont {Martin}\ \bibnamefont
  {Bauer}}, \bibinfo {author} {\bibfnamefont {Matthias}\ \bibnamefont
  {Neubert}}, \bibinfo {author} {\bibfnamefont {Sophie}\ \bibnamefont
  {Renner}}, \bibinfo {author} {\bibfnamefont {Marvin}\ \bibnamefont
  {Schnubel}}, \ and\ \bibinfo {author} {\bibfnamefont {Andrea}\ \bibnamefont
  {Thamm}},\ }\bibfield  {title} {\enquote {\bibinfo {title} {{The Low-Energy
  Effective Theory of Axions and ALPs}},}\ }\href {\doibase
  10.1007/JHEP04(2021)063} {\bibfield  {journal} {\bibinfo  {journal} {JHEP}\
  }\textbf {\bibinfo {volume} {04}},\ \bibinfo {pages} {063} (\bibinfo {year}
  {2021}{\natexlab{b}})},\ \Eprint {http://arxiv.org/abs/2012.12272}
  {arXiv:2012.12272 [hep-ph]} \BibitemShut {NoStop}%
\bibitem [{\citenamefont {Ajimura}\ \emph {et~al.}(2017)\citenamefont {Ajimura}
  \emph {et~al.}}]{Ajimura:2017fld}%
  \BibitemOpen
  \bibfield  {author} {\bibinfo {author} {\bibfnamefont {S.}~\bibnamefont
  {Ajimura}} \emph {et~al.},\ }\bibfield  {title} {\enquote {\bibinfo {title}
  {{Technical Design Report (TDR): Searching for a Sterile Neutrino at J-PARC
  MLF (E56, JSNS2)}},}\ }\href@noop {} {\  (\bibinfo {year} {2017})},\ \Eprint
  {http://arxiv.org/abs/1705.08629} {arXiv:1705.08629 [physics.ins-det]}
  \BibitemShut {NoStop}%
\bibitem [{\citenamefont {Ajimura}\ \emph {et~al.}(2021)\citenamefont {Ajimura}
  \emph {et~al.}}]{JSNS2:2021hyk}%
  \BibitemOpen
  \bibfield  {author} {\bibinfo {author} {\bibfnamefont {S.}~\bibnamefont
  {Ajimura}} \emph {et~al.} (\bibinfo {collaboration} {JSNS2}),\ }\bibfield
  {title} {\enquote {\bibinfo {title} {{The JSNS2 detector}},}\ }\href
  {\doibase 10.1016/j.nima.2021.165742} {\bibfield  {journal} {\bibinfo
  {journal} {Nucl. Instrum. Meth. A}\ }\textbf {\bibinfo {volume} {1014}},\
  \bibinfo {pages} {165742} (\bibinfo {year} {2021})},\ \Eprint
  {http://arxiv.org/abs/2104.13169} {arXiv:2104.13169 [physics.ins-det]}
  \BibitemShut {NoStop}%
\bibitem [{\citenamefont {Maruyama}(2022)}]{Maruyama:2022juu}%
  \BibitemOpen
  \bibfield  {author} {\bibinfo {author} {\bibfnamefont {Takasumi}\
  \bibnamefont {Maruyama}} (\bibinfo {collaboration} {JSNS2, JSNS2 -II,}),\
  }\bibfield  {title} {\enquote {\bibinfo {title} {{The status of JSNS$^2$ and
  JSNS$^2$-II}},}\ }\href {\doibase 10.22323/1.402.0159} {\bibfield  {journal}
  {\bibinfo  {journal} {PoS}\ }\textbf {\bibinfo {volume} {NuFact2021}},\
  \bibinfo {pages} {159} (\bibinfo {year} {2022})}\BibitemShut {NoStop}%
\bibitem [{\citenamefont {Aguilar}\ \emph {et~al.}(2001)\citenamefont {Aguilar}
  \emph {et~al.}}]{LSND:2001aii}%
  \BibitemOpen
  \bibfield  {author} {\bibinfo {author} {\bibfnamefont {A.}~\bibnamefont
  {Aguilar}} \emph {et~al.} (\bibinfo {collaboration} {LSND}),\ }\bibfield
  {title} {\enquote {\bibinfo {title} {{Evidence for neutrino oscillations from
  the observation of $\bar{\nu}_e$ appearance in a $\bar{\nu}_\mu$ beam}},}\
  }\href {\doibase 10.1103/PhysRevD.64.112007} {\bibfield  {journal} {\bibinfo
  {journal} {Phys. Rev. D}\ }\textbf {\bibinfo {volume} {64}},\ \bibinfo
  {pages} {112007} (\bibinfo {year} {2001})},\ \Eprint
  {http://arxiv.org/abs/hep-ex/0104049} {arXiv:hep-ex/0104049} \BibitemShut
  {NoStop}%
\bibitem [{\citenamefont {Gudima}\ \emph {et~al.}(2010)\citenamefont {Gudima},
  \citenamefont {Mokhov},\ and\ \citenamefont {Striganov}}]{Gudima:2009zz}%
  \BibitemOpen
  \bibfield  {author} {\bibinfo {author} {\bibfnamefont {Konstantin~K.}\
  \bibnamefont {Gudima}}, \bibinfo {author} {\bibfnamefont {N.~V.}\
  \bibnamefont {Mokhov}}, \ and\ \bibinfo {author} {\bibfnamefont {S.~I.}\
  \bibnamefont {Striganov}},\ }\bibfield  {title} {\enquote {\bibinfo {title}
  {{Kaon Yields for 2 to 8 GeV Proton Beams}},}\ }in\ \href {\doibase
  10.1142/9789814317290_0012} {\emph {\bibinfo {booktitle} {{Workshop on
  Applications of High Intensity Proton Accelerators}}}}\ (\bibinfo {year}
  {2010})\ pp.\ \bibinfo {pages} {115--119}\BibitemShut {NoStop}%
\bibitem [{\citenamefont {Axani}\ \emph {et~al.}(2015)\citenamefont {Axani},
  \citenamefont {Collin}, \citenamefont {Conrad}, \citenamefont {Shaevitz},
  \citenamefont {Spitz},\ and\ \citenamefont {Wongjirad}}]{Axani:2015dha}%
  \BibitemOpen
  \bibfield  {author} {\bibinfo {author} {\bibfnamefont {S}~\bibnamefont
  {Axani}}, \bibinfo {author} {\bibfnamefont {G}~\bibnamefont {Collin}},
  \bibinfo {author} {\bibfnamefont {JM}~\bibnamefont {Conrad}}, \bibinfo
  {author} {\bibfnamefont {MH}~\bibnamefont {Shaevitz}}, \bibinfo {author}
  {\bibfnamefont {J}~\bibnamefont {Spitz}}, \ and\ \bibinfo {author}
  {\bibfnamefont {T}~\bibnamefont {Wongjirad}},\ }\bibfield  {title} {\enquote
  {\bibinfo {title} {{Decisive disappearance search at high $\Delta m^2$ with
  monoenergetic muon neutrinos}},}\ }\href {\doibase
  10.1103/PhysRevD.92.092010} {\bibfield  {journal} {\bibinfo  {journal} {Phys.
  Rev. D}\ }\textbf {\bibinfo {volume} {92}},\ \bibinfo {pages} {092010}
  (\bibinfo {year} {2015})},\ \Eprint {http://arxiv.org/abs/1506.05811}
  {arXiv:1506.05811 [physics.ins-det]} \BibitemShut {NoStop}%
\bibitem [{\citenamefont {Jordan}\ \emph {et~al.}(2018)\citenamefont {Jordan},
  \citenamefont {Kahn}, \citenamefont {Krnjaic}, \citenamefont {Moschella},\
  and\ \citenamefont {Spitz}}]{Jordan:2018gcd}%
  \BibitemOpen
  \bibfield  {author} {\bibinfo {author} {\bibfnamefont {Johnathon~R.}\
  \bibnamefont {Jordan}}, \bibinfo {author} {\bibfnamefont {Yonatan}\
  \bibnamefont {Kahn}}, \bibinfo {author} {\bibfnamefont {Gordan}\ \bibnamefont
  {Krnjaic}}, \bibinfo {author} {\bibfnamefont {Matthew}\ \bibnamefont
  {Moschella}}, \ and\ \bibinfo {author} {\bibfnamefont {Joshua}\ \bibnamefont
  {Spitz}},\ }\bibfield  {title} {\enquote {\bibinfo {title} {{Signatures of
  Pseudo-Dirac Dark Matter at High-Intensity Neutrino Experiments}},}\ }\href
  {\doibase 10.1103/PhysRevD.98.075020} {\bibfield  {journal} {\bibinfo
  {journal} {Phys. Rev. D}\ }\textbf {\bibinfo {volume} {98}},\ \bibinfo
  {pages} {075020} (\bibinfo {year} {2018})},\ \Eprint
  {http://arxiv.org/abs/1806.05185} {arXiv:1806.05185 [hep-ph]} \BibitemShut
  {NoStop}%
\bibitem [{\citenamefont {Jeon}(2021)}]{hyoungku_jeon_2021_5784148}%
  \BibitemOpen
  \bibfield  {author} {\bibinfo {author} {\bibfnamefont {HyoungKu}\
  \bibnamefont {Jeon}},\ }\href {\doibase 10.5281/zenodo.5784148} {\enquote
  {\bibinfo {title} {{Cosmic ray induced Background study at the JSNS2
  experiment}},}\ } (\bibinfo {year} {2021}),\ \bibinfo {note} {for the JSNS2
  collaboration}\BibitemShut {NoStop}%
\bibitem [{\citenamefont {Ajimura}\ \emph {et~al.}(2020)\citenamefont {Ajimura}
  \emph {et~al.}}]{Ajimura:2020qni}%
  \BibitemOpen
  \bibfield  {author} {\bibinfo {author} {\bibfnamefont {S.}~\bibnamefont
  {Ajimura}} \emph {et~al.},\ }\bibfield  {title} {\enquote {\bibinfo {title}
  {{Proposal: JSNS$^2$-II}},}\ }\href@noop {} {\  (\bibinfo {year} {2020})},\
  \Eprint {http://arxiv.org/abs/2012.10807} {arXiv:2012.10807 [hep-ex]}
  \BibitemShut {NoStop}%
\bibitem [{\citenamefont {Kim}\ \emph {et~al.}(2011)\citenamefont {Kim},
  \citenamefont {Seo},\ and\ \citenamefont {Joo}}]{kim2011measurement}%
  \BibitemOpen
  \bibfield  {author} {\bibinfo {author} {\bibfnamefont {Byoung~Chan}\
  \bibnamefont {Kim}}, \bibinfo {author} {\bibfnamefont {Seung~Won}\
  \bibnamefont {Seo}}, \ and\ \bibinfo {author} {\bibfnamefont {Kyung~Kwang}\
  \bibnamefont {Joo}},\ }\bibfield  {title} {\enquote {\bibinfo {title}
  {Measurement of the density of liquid scintillator solvents for neutrino
  experiments},}\ }\href@noop {} {\  (\bibinfo {year} {2011})}\BibitemShut
  {NoStop}%
\bibitem [{\citenamefont {Kelly}\ and\ \citenamefont
  {Machado}(2021)}]{Kelly:2021xbv}%
  \BibitemOpen
  \bibfield  {author} {\bibinfo {author} {\bibfnamefont {Kevin~James}\
  \bibnamefont {Kelly}}\ and\ \bibinfo {author} {\bibfnamefont {Pedro A.~N.}\
  \bibnamefont {Machado}},\ }\bibfield  {title} {\enquote {\bibinfo {title}
  {{MicroBooNE experiment, NuMI absorber, and heavy neutral leptons}},}\ }\href
  {\doibase 10.1103/PhysRevD.104.055015} {\bibfield  {journal} {\bibinfo
  {journal} {Phys. Rev. D}\ }\textbf {\bibinfo {volume} {104}},\ \bibinfo
  {pages} {055015} (\bibinfo {year} {2021})},\ \Eprint
  {http://arxiv.org/abs/2106.06548} {arXiv:2106.06548 [hep-ph]} \BibitemShut
  {NoStop}%
\bibitem [{\citenamefont {Toups}\ \emph
  {et~al.}(2022{\natexlab{a}})\citenamefont {Toups} \emph
  {et~al.}}]{Toups:2022yxs}%
  \BibitemOpen
  \bibfield  {author} {\bibinfo {author} {\bibfnamefont {M.}~\bibnamefont
  {Toups}} \emph {et~al.},\ }\bibfield  {title} {\enquote {\bibinfo {title}
  {{PIP2-BD: GeV Proton Beam Dump at Fermilab's PIP-II Linac}},}\ }in\
  \href@noop {} {\emph {\bibinfo {booktitle} {{Snowmass 2021}}}}\ (\bibinfo
  {year} {2022})\ \Eprint {http://arxiv.org/abs/2203.08079} {arXiv:2203.08079
  [hep-ex]} \BibitemShut {NoStop}%
\bibitem [{\citenamefont {Aguilar-Arevalo}\ \emph {et~al.}(2022)\citenamefont
  {Aguilar-Arevalo} \emph {et~al.}}]{CCM:2021yzc}%
  \BibitemOpen
  \bibfield  {author} {\bibinfo {author} {\bibfnamefont {A.~A.}\ \bibnamefont
  {Aguilar-Arevalo}} \emph {et~al.} (\bibinfo {collaboration} {CCM}),\
  }\bibfield  {title} {\enquote {\bibinfo {title} {{First Leptophobic Dark
  Matter Search from the Coherent\textendash{}CAPTAIN-Mills Liquid Argon
  Detector}},}\ }\href {\doibase 10.1103/PhysRevLett.129.021801} {\bibfield
  {journal} {\bibinfo  {journal} {Phys. Rev. Lett.}\ }\textbf {\bibinfo
  {volume} {129}},\ \bibinfo {pages} {021801} (\bibinfo {year} {2022})},\
  \Eprint {http://arxiv.org/abs/2109.14146} {arXiv:2109.14146 [hep-ex]}
  \BibitemShut {NoStop}%
\bibitem [{\citenamefont {Akimov}\ \emph {et~al.}(2020)\citenamefont {Akimov}
  \emph {et~al.}}]{COHERENT:2019kwz}%
  \BibitemOpen
  \bibfield  {author} {\bibinfo {author} {\bibfnamefont {D.}~\bibnamefont
  {Akimov}} \emph {et~al.} (\bibinfo {collaboration} {COHERENT}),\ }\bibfield
  {title} {\enquote {\bibinfo {title} {{Sensitivity of the COHERENT Experiment
  to Accelerator-Produced Dark Matter}},}\ }\href {\doibase
  10.1103/PhysRevD.102.052007} {\bibfield  {journal} {\bibinfo  {journal}
  {Phys. Rev. D}\ }\textbf {\bibinfo {volume} {102}},\ \bibinfo {pages}
  {052007} (\bibinfo {year} {2020})},\ \Eprint
  {http://arxiv.org/abs/1911.06422} {arXiv:1911.06422 [hep-ex]} \BibitemShut
  {NoStop}%
\bibitem [{\citenamefont {Toups}\ \emph
  {et~al.}(2022{\natexlab{b}})\citenamefont {Toups} \emph
  {et~al.}}]{Toups:2022knq}%
  \BibitemOpen
  \bibfield  {author} {\bibinfo {author} {\bibfnamefont {Matt}\ \bibnamefont
  {Toups}} \emph {et~al.},\ }\bibfield  {title} {\enquote {\bibinfo {title}
  {{SBN-BD: $\mathcal{O}$(10 GeV) Proton Beam Dump at Fermilab's PIP-II
  Linac}},}\ }in\ \href@noop {} {\emph {\bibinfo {booktitle} {{Snowmass
  2021}}}}\ (\bibinfo {year} {2022})\ \Eprint {http://arxiv.org/abs/2203.08102}
  {arXiv:2203.08102 [hep-ex]} \BibitemShut {NoStop}%
\bibitem [{Matheus Hostert()}]{Matheus_Private}%
  \BibitemOpen
  Matheus Hostert,\ \href@noop {} {} (\bibinfo {year} {2023}),\ \bibinfo {note}
  {private communication}\BibitemShut {NoStop}%
\end{thebibliography}%
\end{document}